\newcommand\papertitle{High precision modeling of polarized signals: Moment expansion method generalized to spin-2 fields}

\documentclass[twocolumn]{aa}

\makeatletter
\renewcommand*\aa@pageof{, page \thepage{} of \pageref*{LastPage}}
\makeatother
\usepackage{bbm}
\usepackage{graphicx}
\usepackage{txfonts}
\usepackage{url}
\usepackage{natbib}
\bibpunct{(}{)}{;}{a}{}{,} 
\usepackage{color}
\usepackage{colortbl} 
\usepackage[dvipsnames]{xcolor}
\usepackage{enumitem}
\usepackage[small]{caption}
\usepackage{tipa}

\usepackage[switch]{lineno}
\usepackage[breaklinks, colorlinks, citecolor=blue]{hyperref}
\usepackage{subfigure}
\usepackage{epsfig}
\usepackage{multirow}
\usepackage{array}

\usepackage{url}
\usepackage{psfrag}
\usepackage[T1]{fontenc}
\usepackage{rotating}
\usepackage{soul}
\usepackage{wrapfig}
\usepackage{tikz}

\makeatletter
\def\env@cases{
  \let\@ifnextchar\new@ifnextchar
  \left\lbrace
  \def\arraystretch{1.2}
  \array{l@{}l@{}}
}
\makeatother

\definecolor{mygreen}{RGB}{104,198,107}
\definecolor{myred}{RGB}{252,137,125}
\definecolor{myyellow}{RGB}{252,225,126}
\definecolor{mygrey}{RGB}{215,215,215}

\def\fracnu{\left(\frac{\nu}{\nu_0}\right)}
\def\lnnu{\ln\fracnu}
\def\i{\mathbbm{i}}
\def\barp{\bar{p}}
\def\barpv{\vec{\bar{p}}}

\def\vecn{\hat{\vec{n}}}
\def\lb{\textit{LiteBIRD}}
\def\wmap{\textit{WMAP}}
\def\planck{\textit{Planck}}
\def\so{{The Simons Observatory}}
\def\Iunit{~{\rm Jy/sr}}

\newcommand{\expf}[1]{{\rm e}^{#1}}
\newcommand{\gammab}{\bar{\gamma}}
\def\barppol{\bar{p}}
\def\barppolv{\vec{\bar{p}}}
\def\barA{\bar{A}}
\def\betaGB{\beta_{\rm GB}}
\def\betaGBb{\bar{\beta}_{\rm GB}}
\def\betad{\beta_{\rm d}}
\def\betadb{\bar{\beta}_{\rm d}}
\def\spinpol{\mathcal{P}}
\def\hatpol{\hat{P}}

\def\polnu{P_\nu}
\def\prob{\mathbb{P}}

\begin{document}

\title{\papertitle}

\offprints{\url{leo.vacher@irap.omp.eu}}
\authorrunning{Vacher et al.}
\titlerunning{High precision modeling of polarized signals}

\author{L. Vacher$^1$ 
\and J. Chluba$^2$ 
\and J. Aumont$^1$
\and A. Rotti$^2$   
\and L. Montier$^1$   
}
\institute{
$^1$ IRAP, Universit\'e de Toulouse, CNRS, CNES, UPS, Toulouse, France\label{inst1}
\\
$^2$ Jodrell Bank Centre for Astrophysics, Alan Turing Building, University of Manchester, Manchester M13 9PL\label{inst2}
}

\abstract{
The modeling and removal of foregrounds poses a major challenge to searches for signals from inflation using the cosmic microwave background (CMB). In particular, the modeling of CMB foregrounds including various spatial averaging effects introduces multiple complications that will have to be accounted for in upcoming analyses.
In this work, we introduce the generalization of the intensity moment expansion to the spin-2 field of linear polarization: the spin-moment expansion.
Within this framework, moments become spin-2 objects that are directly related to the underlying spectral parameters and polarization angle distribution functions. In obtaining the required expressions for the polarization modeling, we highlight the similarities and differences with the intensity moment methods. A spinor rotation in the complex plane with frequency naturally arises from the first order moment when the signal contains both spectral parameters and polarization angle variations. Additional dependencies are introduced at higher order, and we demonstrate how these can be accounted with several illustrative examples.
Our new modeling of the polarized signals reveals to be a powerful tool to model the frequency dependence of the polarization angle. As such, it can be immediately applied to numerous astrophysical situations.}

\keywords{Cosmology, CMB, Foregrounds, Interstellar medium}

\maketitle

\section{Introduction}
%---------------------------------
A significant international effort has been undertaken to deploy large-scale surveys of the comsic microwave background (CMB) signal in the present, near, and far future. Multiple telescopes, such as {\sc ACT} \citep{ACT}, {\sc SPT} \citep{SPT}, \so{} \citep{SimonsObservatory}, and {\sc CMB-S4} \citep{CMBS4}, are or will be observing large portions of the sky from the ground. Similarly, from space we are eagerly awaiting \lb{} \citep{Ptep}, and in the future possibly even more ambitious CMB imagers \citep{PICO, Voyage2050}. In addition, we can hope for a CMB spectrometer such as {\it PIXIE} \citep{PIXIE}
to target CMB spectral distortions \citep{ChlubaVoy2050}. 
The scientific targets are manyfold and of prime importance to cosmology, astrophysics, and high-energy physics.

Ever-increasing instrumental sensitivities imply that one also becomes sensitive to faint and complex effects that need to be properly modeled. 
%-
The fine characterization of polarized astrophysical signals thus becomes an increasingly complicated and important challenge to cosmological analyses.
%-
The stakes are twofold: first, we wish to reach a better understanding of the sources themselves and the complex physics at play in the emission. This is of importance for Galactic physics, physics of recombination at the last scattering surface with the primordial anisotropies of the cosmic microwave background (CMB) and their secondary sources from the physics of galaxy clusters with the Sunyaev-Zeldovich (SZ) effect \citep{CosmoSZ, Mroczkowski2019}, or lensing \citep{Lensing}.
Secondly, high-precision component separation is required to remove the diffuse foregrounds and access the faint perturbations of the CMB polarized signal. Finding how to deal with them is an unavoidable step in addressing questions that concern the cosmic history and high-energy physics through inferring key cosmological parameters as the reionization depth $\tau$ \citep{ReioWise} or the scalar-to-tensor ratio $r$, which probes inflation \citep{inflationhist1,inflationhist2,inflationhist3}. 
%-
The philosophy of the present work is that one cannot properly deal with the second point without a detailed understanding of the polarized Galactic emission, deeply rooted in the physics. 

%Intensity%%%%%%
It is known that the emission properties of the diffuse interstellar medium (ISM) change across the Galaxy on small and large scales. This assertion is well motivated from the Galactic magnetic field physics and the distribution of dust grain shapes and composition  \citep{Ferriere,Jaffe2013}, which is supported by numerous observations \citep{Ysard2013,vardustdisk,vardustdisk2,PlanckL,PlanckDust2,pelgrims2021}.
%-
In observational conditions, averages over Galactic voxels with different spectral parameters are then unavoidable. They occur in several situations: along the line of sight inside the Galaxy, which cannot be reduced or avoided with instrumental considerations; between lines of sight inside the instrumental beam; and over patches of the sky when doing a spherical harmonic decomposition of the signal or averaging the data otherwise.
%-
Two related consequences follow immediately for the signal in intensity: if the fundamental spectral energy distributions (SEDs) are nonlinear, the average SED of the total signal differs from the canonical SED of the voxel. We refer to this phenomenon as SED distortions. The SED is distorted differently from one point of the sky to another, breaking the correlation between the maps at different frequency bands, leading to inaccurate extrapolations from one to another: we refer to this phenomenon as frequency decorrelation \citep[see e.g.,][]{tassis,PlanckL,pelgrims2021}. 
%-

To treat these averages in connection with CMB foregrounds, the (Taylor) moment expansion formalism was proposed \citep{Chluba}. A similar idea had been applied to the modeling of Sunyaev-Zeldovich signals, showing how spatial and frequency information can be nicely separated \citep{Chluba2012SZpack}.
The moment formalism has proven to be very powerful when applied to component separation at the map level \citep{MILC,Remazeilles_etal_2016,RemazeillesmomentsILC}. 
%-
A straightforward generalization to harmonic space and cross-frequency power-spectra domain has also proven to be useful \citep{Mangilli,Azzoni,Vacher21}. 

%Polar-----
%
While the original formulation of the moment method was focused on the intensity, it was stressed that an extension to polarization can be readily obtained \citep{Chluba}, which is what we intend to do in the present work.
Indeed, several applications already used the moment expansion method for polarized signals, treating the $B$-mode signal as an intensity \citep[e.g.,][]{RemazeillesmomentsILC,Azzoni,Vacher21}.  %disregarding the unique geometrical properties of polarization
One can also find a similar approach in the Delta-map method \citep{moment_polar}, which used first order terms of the $Q/U$-intensity moments and already suggested a common treatment for the pair $(Q,U)$.

In this work, we plan to rigorously derive and extend the moment expansion method to polarized signals. To do so, the Stokes parameters must be treated together, as the components of a single complex object.
In the most general cases, extra subtleties come into play, which were not captured or discussed before. There are numerous advantages from thinking of linear polarization as a spin-2 quantity. As such they are not only described by a scalar quantity but also by an angle: the polarization angle.
%-
The averaging processes listed above will have one extra consequence for polarized signal: additionally to the spectral parameters, multiple angles will be mixed along and between lines of sight. We refer to this phenomenon as polarized mixing. In the presence of polarized mixing, the total signal will exhibit a frequency-dependent polarization angle. Being able to accurately model this frequency-dependent rotation from physically-motivated considerations represents a thorny challenge. 
%-
In this work, we attempt to provide this extension to linearly polarized signal in a formal, natural, complete and self-consistent way. 
We pay particular attention to the formulation in terms of SED parameter distribution functions, which really is the origin of the name "moment expansion". We subsequently see that this rewriting offers a powerful framework to grasp polarized mixing and its consequences. 
%\reviewcorrect{Surprising new results arise along the way, as the correction of the pivot spectral parameter, which becomes a complex number, provides an analytical expression for the frequency dependence of the polarization angle in the perturbative regime. }

After a review of the intensity moment formalism in Sec.~\ref{sec:momintensity}, we discuss the nature of linear polarization and introduce the spin-moment formalism in a single line of sight in Sec.~\ref{sec:spin-moments}. In Sec.~\ref{sec:applications}, we explore different example of sums of canonical SEDs along a line of sight, that are of astrophysical relevance. We study them both analytically and through a fitting procedure, demonstrating the ability of the spin-moment formalism to grasp distortions of the polarized SED.
%we analytically explore several examples of astrophysical relevance build of discrete sums of SEDs along a single line-of-sight and perform complex curves fitting, demonstrating the ability of this new formalism to grasp distortions of the polarized SED, even in extreme cases.
%
In Sec.~\ref{sec:applications2}, we generalize the formalism to deal with other kind of averaging effects: spherical harmonic transforms and instrumental effects. 
In Sec.~\ref{sec:discussion}, we discuss cases with extra complexity as Faraday rotation and more general voxel SEDs.
%we draw a state of the art of how this method is related to previous works. 
Finally, we conclude in Sec.~\ref{sec:conclusion}.    

\vspace{-4mm}
\section{Intensity moment expansion}% along a single line-of-sight} 
\label{sec:momintensity}
%--------------------------------------
Before we discuss the generalization of the moment expansion for polarized light, we briefly recall the logical steps followed in \cite{Chluba} to obtain the moment expansion in intensity. For now, we neglect beam averaging effects or expansions into spherical harmonic, but we cover these in Sec.~\ref{sec:applications2}. 

We start by considering various voxels along a line of sight in the direction $\vec{\hat{n}}$, which is described by an affine parameter $s$. Every voxel emits with an SED\footnote{Hereafter, we use the shorthand notation for frequency dependent quantities $X_\nu\equiv X(\nu)$.}:
%--------------------------------------
\begin{align}
I_{\nu}(A(s),\vec{p}(s))=A(s)\,\hat{I}_\nu(\vec{p}(s)),
\label{eq:I_nu}
\end{align}
%--------------------------------------
where $\hat{I}_\nu$ is referred to as the fundamental SED with $N$ spectral parameters, $\vec{p}(s) = \{p_{1}(s),p_{2}(s),\dots,p_{N}(s)\}$. The amplitude or weight parameter, $A(s)$, determines the relative contribution of each voxel to the total intensity.\footnote{One simple example is the power law: $I_\nu(A,\beta)=A\,(\nu/\nu_0)^\beta$ where $\vec{p}=\{\beta\}$ has dimension one ($N=1$) and $\hat{I}_\nu(\beta)=(\nu/\nu_0)^\beta$. In this, $\nu_0$ is arbitrarily defined, but in practical applications the choice is normally data-driven (e.g. motivated by the location of sensitive bands in
experiments such as for Planck or WMAP) and $A$ is the overall weight, with dimension that depends on the situation (see Appendix~\ref{sec:weights}).} 
The resulting total SED is given by an average along the line of sight, which we shall denote by $\langle \dots \rangle$. This average can be explicitly written in terms of an integral over the affine parameters, $s$, or, alternatively, as an integral of the intensity over the spectral parameter distribution function in the direction $\vecn$ \citep[e.g.,][]{Chluba, MILC}:
%--------------------------------------
\begin{align}
\label{eq:I_av_los}
\langle I_\nu(A\,,\vec{p})\rangle
&=\int \frac{{\rm d}A(s)}{{\rm d}s}\,\hat{I}_\nu(\vec{p}(s)) \,{\rm d}s
\equiv\int \prob(\vec{p}, \vecn) \,\hat{I}_\nu(\vec{p}) \,{\rm d}^N p. 
%\langle I_\nu(A\,,\vec{p})\rangle
%&=\sum_k A_k\hat{I}_\nu(\vec{p}_k)
%\equiv\int \prob(\vec{p}, \vecn) \,\hat{I}_\nu(\vec{p}) \,{\rm d}^N p. 
\end{align}
%--------------------------------------
In the second definition, we introduced the distribution $\prob(\vec{p}, \vecn)$ of the spectral parameters, $\vec{p}$, along the fixed line of sight $\vecn$, with the relative weights absorbed into the distribution itself.
We note that the distribution $\prob(\vec{p}, \vecn)$ is not necessarily normalized to unity, as it determines the relative weight of each SED shape to the total.
For convenience we shall define the average amplitude parameter as $\barA=\langle A\rangle=\int [{\rm d}A(s)/{\rm d}s]\,{\rm d}s=\int \prob(\vec{p}, \vecn) \,{\rm d}^N p$.

In order to provide a perturbative model of the average SED, the spectral dependence $\hat{I}_\nu$ in Eq.~\eqref{eq:I_av_los} can be expanded into a Taylor series with respect to $\vec{p}$ around the pivot $\barpv$ as:
%--------------------------------------
\begin{align}
\label{eq:taylorexp}
     \hat{I}_\nu(\vec{p}) &= \hat{I}_\nu(\barpv) 
    + \sum_j (p_j-\barp_j)\,
    \partial_{\barp_j} \hat{I}_\nu(\barpv)
    \nonumber \\
    &\quad+ \frac{1}{2}\sum_{j,k} (p_j-\barp_j)(p_k-\barp_k)\,
    \partial_{\barp_j}\partial_{\barp_k}\hat{I}_\nu(\barpv)
    \\ \nonumber 
    &\qquad+ \frac{1}{3!}\sum_{j,k,l}(p_j-\barp_j)(p_k-\barp_k)(p_l-\barp_l)\,
    \partial_{\barp_j}\partial_{\barp_k}\partial_{\barp_l}\hat{I}_\nu(\barpv) 
    \\ \nonumber 
    &\qquad\quad+ \dots \,.
\end{align}
%--------------------------------------
Here, we used the shorthand notation $\partial_{\barp_j} X(\barpv) \equiv \partial X(\barpv)/\partial \barp_j$. The pivot value $\bar{\vec{p}}$ around which the series is carried out can be fixed by asking for the first term of the expansion to vanish upon averaging: $\langle A \sum_j (p_j-\barp_j)\rangle=0$. This minimizes the required terms in the Taylor series and leads to: 
%--------------------------------------
\begin{equation}
\label{eq:pivot}
    \barpv
    = \frac{\langle A\vec{p}\rangle}{\barA}
    = \frac{\int \prob(\vec{p}, \vecn) \,\vec{p}\,{\rm d}^N p}{\int \prob(\vec{p}, \vecn) \,{\rm d}^N p}.
\end{equation}
%--------------------------------------
Next we introduce the moment coefficients of order $\alpha$
%--------------------------------------
\begin{align}
\label{eq:moments}
\omega_\alpha^{p_j\dots p_l}
&=\frac{\langle A\,(p_j-\barp_j)\dots(p_l-\barp_l)\rangle}{\barA}
\nonumber\\
&=\frac{\int \prob(\vec{p}, \vecn) \,(p_j-\barp_j)\dots(p_l-\barp_l)\,{\rm d}^N p}{\int \prob(\vec{p}, \vecn) \,{\rm d}^N p},
\end{align}
%--------------------------------------
with $\alpha$ being the number of parameters over which the average is done and the maximal order of the derivative associated with the moment coefficient. One can then write the total intensity as an expansion in terms of these moments:
%--------------------------------------
\begin{align}
\label{eq:moments-intensity}
    \langle I_\nu(A,\vec{p})\rangle &=  I_\nu(\barA,\barpv) + \sum_j^N\omega_1^{p_j} \partial_{\barp_j} I_\nu(\barA,\barpv)
    \nonumber \\
    &\quad+ 
    \frac{1}{2}\sum_{j,k}^N\omega_2^{p_j p_k}\partial_{\barp_j}\partial_{\barp_k}I_\nu(\barA,\barpv)
    \nonumber \\
    &\qquad+ \frac{1}{3!}\sum_{j,k,l}^N\omega_3^{p_j p_k p_l}\partial_{\barp_j}\partial_{\barp_k}\partial_{\barp_l}I_\nu(\barA,\barpv) + \dots \,.
\end{align}
%--------------------------------------
Here, all the $\omega^{p_i}_1$ are zero when using the value for $\barpv$ as given by Eq.~\eqref{eq:pivot}, while higher order moments capture the complexities added by line-of-sight averaging effects. In applications, the pivot value can be obtained using an iterative process by starting with a reasonable guess for $\barpv$ and then correcting the solution by the values of $\omega^{p_i}_1\equiv \Delta p_i$ using $\bar{p}_j\rightarrow \bar{p}'_j= \bar{p}_{j} + \Delta p_j$. This iterative process assumes that the moments are perturbative and convergence can be obtained with a finite number of terms.

\vspace{-3mm}
\section{Moment expansion of polarized signals}% along a single line-of-sight}
\label{sec:spin-moments}
%-------------------------------------
In this section, we generalize the intensity moment expansion to polarization. We start by summarizing a few general aspects about how to describe polarized light and then highlight some of the important differences between intensity and polarization.

\vspace{-3mm}
\subsection{General introduction to polarized SED}
\label{sec:intro-polar}
%-------------------------------------
A polarized signal is fully described by the four real Stokes parameters $I_\nu,Q_\nu,U_\nu,V_\nu$, all of them being frequency dependent quantities. 
%-------------------------------------
As before, $I_\nu$ describes the total (i.e., unpolarized + polarized) intensity, while the pair $(Q_\nu,U_\nu)$ and $V_\nu$ respectively quantify the linearly and the circularly polarized part of the photon field. Both $I_\nu$ and $V_\nu$ are scalar fields meaning that they are invariant quantities under transformations of the frame in which they are evaluated. As such, they can be described using the intensity moment expansion of Sec.~\ref{sec:momintensity}. However a general treatment of polarized light including $V_\nu$ with the moment expansion could introduce extra subtleties which go beyond the scope of this work. Henceforth, we assume $V_\nu=0. $\footnote{This is justified for CMB signals and component separation since classical physics in the primordial plasma is not expected to be source of any significant circular polarization \citep{Hirata2018Circ, Inomata2019Circ}. Note however that a faint primordial $V$ signal is expected in some models, see e.g. \cite{CMBVmodes}.}

On the other hand, the $Q_\nu$ and $U_\nu$ are coordinate-dependent quantities transforming under frame rotations as the components of a spin-2 object. Therefore, they can be more naturally combined into a single spinor field $\spinpol_\nu$:
%--------------------------------------
\begin{align}
 \spinpol_\nu &= Q_\nu + \i U_\nu = \polnu \expf{2\i\gamma_\nu}, \quad {\rm with} \,\i = \sqrt{-1}.
\end{align}
%--------------------------------------
The spinor's modulus, $\polnu$, is a real positive function called the linear {\it polarization intensity} and its argument defines the {\it polarization angle} $\gamma_\nu$:
%--------------------------------------
\begin{subequations}
\begin{align}
|\spinpol_\nu|&= \polnu= \sqrt{Q_\nu^2 + U_\nu^2}
\\
{\rm arg}(\spinpol_\nu) &= 2\gamma_\nu= \tan^{-1}\left(U_\nu/Q_\nu\right).
\end{align}
\end{subequations}
%--------------------------------------
As a spin-2 quantity, when the frame in which $Q_\nu$ and $U_\nu$ are defined (e.g., by modifying the directions of the polarizers) is rotated by a right handed rotation around the $\vecn$ direction by an angle $\theta$, $\spinpol_\nu$ transforms as \cite{Zaldarriaga1997}:
%--------------------------------------
\begin{equation}
    (\spinpol_\nu)^\prime = \expf{-2\i\theta}\,\spinpol_\nu.
\end{equation}
%--------------------------------------
%
Note that, unless stated otherwise, henceforth we use calligraphic variables for complex quantities (e.g., $\spinpol_\nu, \mathcal{W}, \dots$) and italic font for real quantities (e.g., $\polnu, A, Q_\nu, U_\nu, \omega, \dots$).

%--------------------------------------

\subsection{Origin of frequency-dependent polarization angle from polarized mixing}
%-------------------------------------
The canonical SEDs usually considered in astrophysics (e.g. power laws, blackbodies, gray-bodies, modified blackbodies) generally assume a constant value for $\gamma$, independent of the frequency. This behavior is motivated by the existence of a preferred direction in the physical mechanisms at the origin of polarized emission such as magnetic fields and dust grain shape.
A single emitting voxel in the Galaxy is thus expected to emit with a constant polarization angle as a function of frequency. 

\begin{figure}[ht!]
    \centering
    \includegraphics[width=0.94\columnwidth]{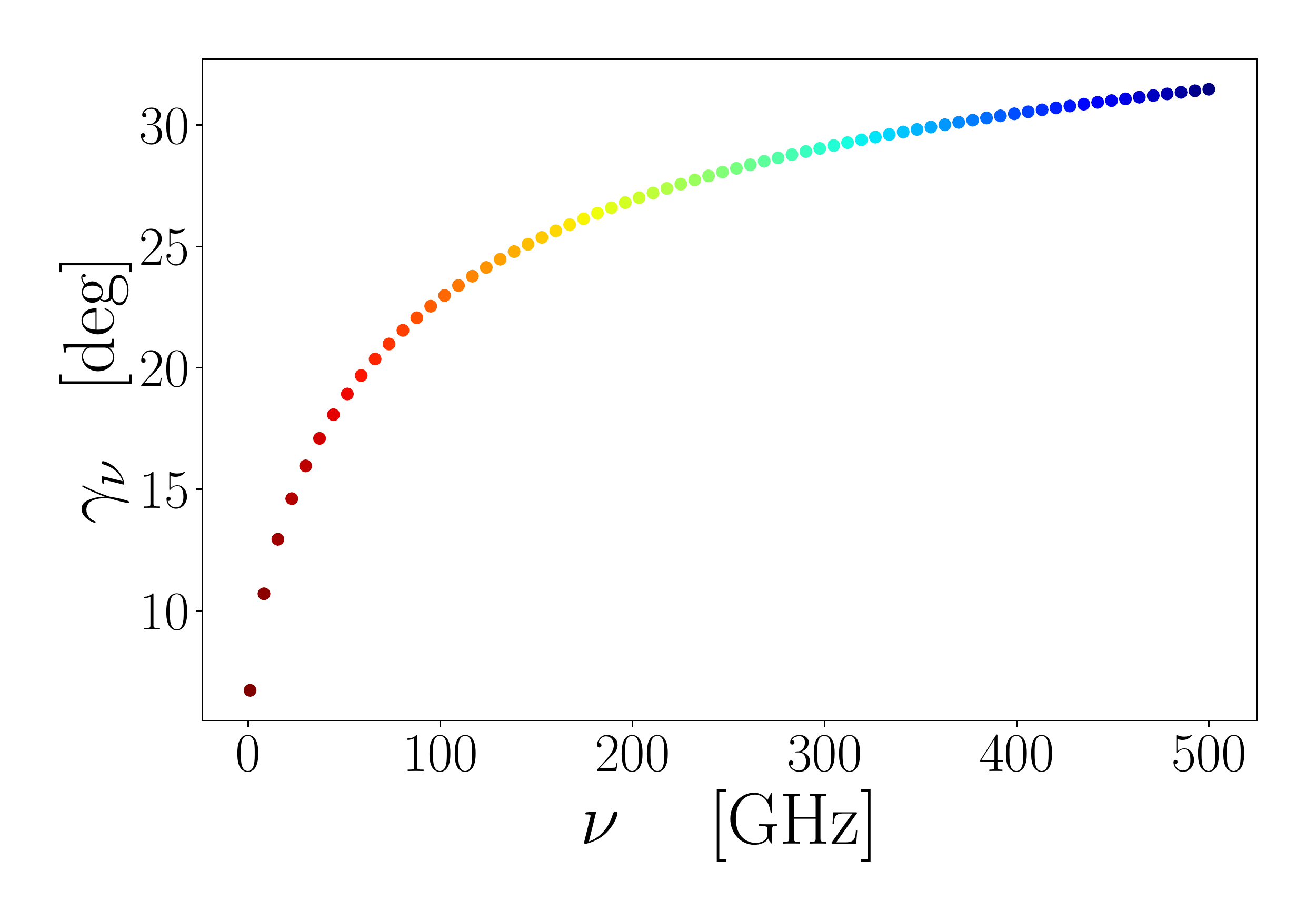}
    \\[-2mm]
    \includegraphics[width=0.9\columnwidth, trim={30mm 0mm 30mm 0},clip]{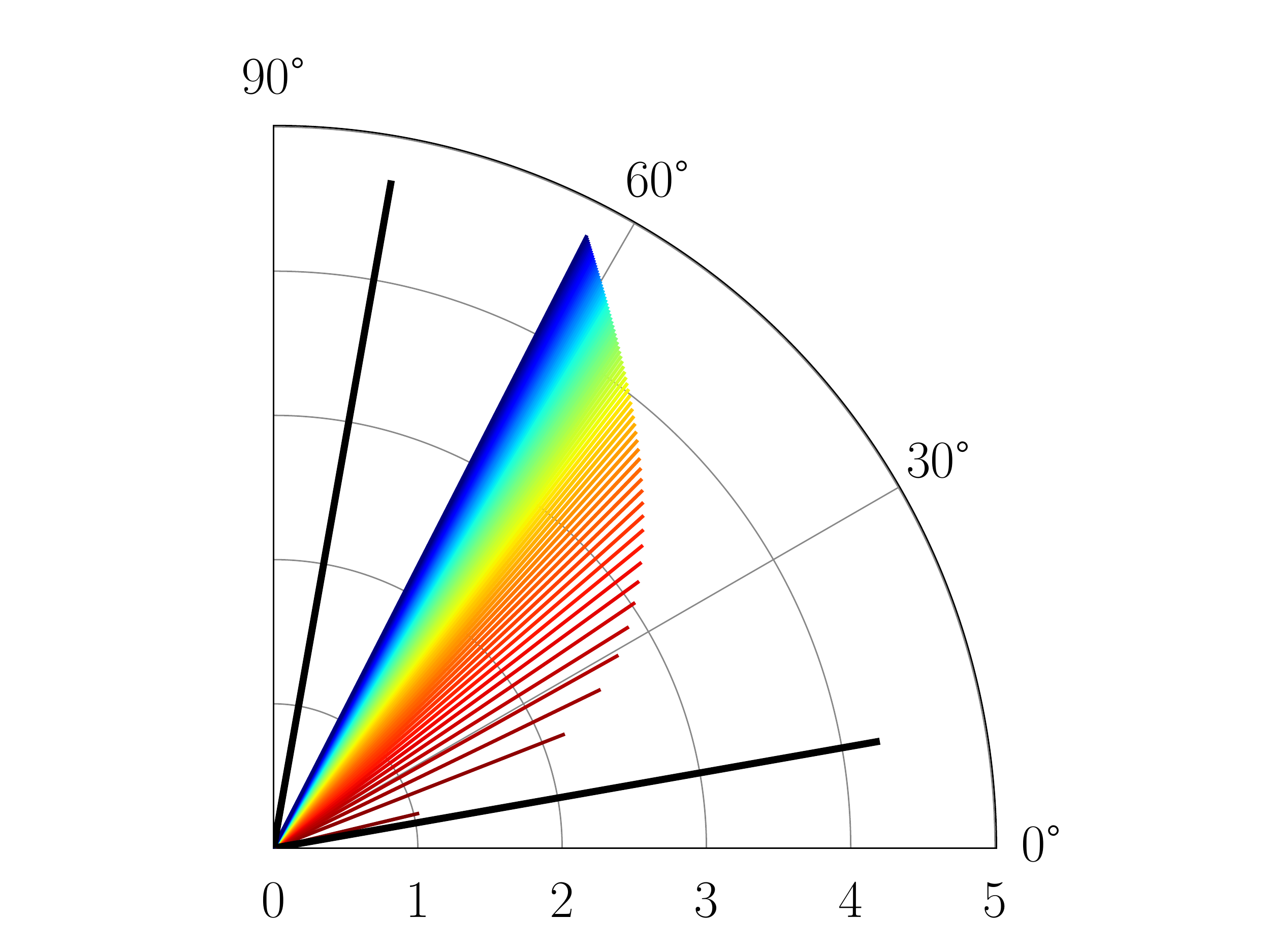}
    \\
    \caption{Illustration of the polarization angle spectral dependence from polarized mixing. {\it Upper panel}: Polarization angle $\gamma_\nu$ as a function of frequency (in GHz) for the sum of two power laws with parameters $A_1=2\Iunit$, $A_2=1\Iunit$, $\beta_1=1.8$, $\beta_2=1.2$, $2\gamma_1=10^{\circ}$ and $2\gamma_2= 80^{\circ}$. The color labels the frequency between $1$\,GHz (dark red) and $500$\,GHz (dark blue). 
    {\it Lower panel}: Polarization spinor $\spinpol_\nu$ (in $\Iunit$) in the complex $(Q,U)$-plane, in the same configuration and with the same color coding. The phase of the spinor is $2\gamma_\nu$. The length of the colored bars and the lines of constant radius represent values of $\log_{10}(\polnu)$. The two black lines represent the two power laws $A_1(\nu/\nu_0)^{\beta_1} \expf{2\i\gamma_1}$ and $A_2(\nu/\nu_0)^{\beta_2}\expf{2\i\gamma_2}$ at 500\,GHz.
 }%the original power-laws of constant arguments at $500$\,GHz before summation.}
    \label{fig:rot_nu}
\end{figure}

Adding voxels with the same polarization angle but varying spectral parameters simply leads to spectral complexity, very much like for intensity. Mixing varying polarization states with the same SED simply leads to a change in the direction of the total polarization, but no extra spectral complexity.
However, if one mixes various SEDs with different polarization angles {\it and} various spectral parameters, the resulting polarized signal will inherit a distorted SED $\polnu'\neq \polnu$ and a frequency dependent $\gamma_\nu$.
%--------------------------------------
\begin{align}
    \polnu(A_1,\vec{p_1})\,\expf{2\i\gamma_1} + \polnu(A_2,\vec{p_2})\,\expf{2\i\gamma_2} + \dots = \polnu'\,\expf{2\i\gamma_\nu}.
\label{eq:sum_power_laws}
\end{align}
%--------------------------------------
This consequence of polarized mixing is illustrated in Fig.~\ref{fig:rot_nu} for a sum of two power laws $A_i (\nu/\nu_0)^{\beta_i} \expf{2\i\gamma_i}$ with $\nu_0=300$\,GHz, $A_1=2\Iunit$, $A_2=1\Iunit$, $\beta_1=1.8$, $\beta_2=1.2$, $2\gamma_1=10^{\circ}$ and $2\gamma_2= 80^{\circ}$. One clearly sees that the resulting spinor $\spinpol_\nu$ rotates in the complex plane with frequency.
%--------------------------------------
Modeling both the distorted SED $P'$ and the frequency dependence of $\gamma_\nu$ in a physically-motivated fashion is nontrivial, but can be achieved when generalizing the moment expansion to polarization.

\vspace{-3mm}
\subsection{Understanding the link to intensity moment expansion}
\label{sec:comments_pol}
%-------------------------------------
To generalize the intensity moment expansion to polarization, we have to discuss how $Q_\nu$ and $U_\nu$ are obtained, and linked to intensity. To characterize the polarization state of the photon field, we measure the intensity with linear polarizers in four directions, $I_{\nu,\parallel}$ and $I_{\nu,\perp}$, which are orthogonal to each other, and $I_{\nu,\times}$ and $I_{\nu,\otimes}$, which are also orthogonal to each other but rotated by $45^\circ$ relative to the previous system. The total intensity $I_\nu$ (polarized $+$ unpolarized), and Stokes $Q_\nu$ and $U_\nu$ are then given by
%--------------------------------------
\begin{subequations}
\begin{align}
(I^{\rm tot}_\nu)^2&=(I^{\rm unpol}_\nu)^2 +
Q_\nu^2+U_\nu^2 
\\
Q_\nu&=\frac{I_{\nu,\parallel}-I_{\nu,\perp}}{2}, \quad U_\nu=\frac{I_{\nu,\times}-I_{\nu,\otimes}}{2}.
\end{align}
\end{subequations}
%--------------------------------------
This shows that both $Q_\nu$ and $U_\nu$ describe {\it differences} between intensities and as such can have positive and negative contributions, depending on which polarizer response dominates. Thinking of each of the intensities $I_{\nu,\parallel}$, $I_{\nu,\perp}$, $I_{\nu,\times}$ and $I_{\nu,\otimes}$ as the cummulative signal from various emitters, means that one can create net polarization by i) varying the number of emitters, that is to say the weight parameter $A$, and ii) changing the spectra of the emitters in the different directions.

%--------------------------------------

To give an example, let us assume that in all directions we have a simple gray-body SED, $I^{\rm GB}_\nu=A\,B_\nu(T)$, where $B_\nu(T)$ is a blackbody spectrum. The linearly polarized radiation of a single voxel is then given by
%--------------------------------------
\begin{align}
\spinpol^{\rm GB}_\nu 
&=\frac{A_\parallel\,B_\nu(T_\parallel)-A_\perp\,B_\nu(T_\perp)}{2}
+\i\frac{A_\times\,B_\nu(T_\times)-A_\otimes\,B_\nu(T_\otimes)}{2}.
\end{align}
%--------------------------------------
Starting with this voxel SED renders the problem quite complicated. For example, just considering the SED of $Q_\nu$, we can write
%--------------------------------------
\begin{align}
\frac{A_\parallel\,B_\nu(T_\parallel)-A_\perp\,B_\nu(T_\perp)}{2}
&=\frac{(A_\parallel-A_\perp)}{2}
\,\frac{B_\nu(T_\parallel)+B_\nu(T_\perp)}{2}
\nonumber\\
&\qquad+\frac{A_\parallel+A_\perp}{2}
\,\frac{B_\nu(T_\parallel)-B_\nu(T_\perp)}{2}.
\end{align}
%--------------------------------------
This means that {\it two} fundamental SED shapes are required: the sum and difference of two blackbody spectra. These are generally not blackbody spectra again \citep{Chluba2004SB} and the moment expansion requires two series. If the number of spectral parameters is extended (here it was only $T$), then the number of fundamental voxel spectra increases rapidly, which can quickly make the situation quite complicated. 

In astrophysical applications, it is commonly assumed that the only source of polarization is through variations of the number of emitters (i.e., the weight parameter $A$) at fixed SED parameters. In our example, this means $T_\parallel\approx T_\perp\approx T_\times\approx T_\otimes=T$, such that $Q_\nu\approx \frac{(A_\parallel-A_\perp)}{2}\,B_\nu(T)$. In this case, the fundamental voxel SED is given by $\hatpol^{\rm GB}_\nu \approx B_\nu(T)$, such that the single polarization state can be characterized by $\spinpol^{\rm GB}_\nu\approx A\,\expf{2\i\gamma} B_\nu(T)$. For the foreground examples treated below, we similarly assume that inside a given voxel the spectral parameters remain constants. We further discuss how to go beyond this assumption in Sec.~\ref{sec:SEDvoxelvar}.

\vspace{-3mm}
\subsection{Spin moments: Moment expansion for spin-2 quantities}
%--------------------------------------
We now generalize the moment expansion for intensity presented in Sec.~\ref{sec:momintensity} to polarized signals. We discuss how this new framework arises naturally from the previous one and provides a powerful tool allowing us to model the frequency dependence of $\gamma_\nu$ in the presence of polarized mixing.

The generalization is indeed quite straightforward. For the intensity moment expansion, we performed a Taylor series in the spectral parameters for each emitting volume element, in Eq.~\eqref{eq:taylorexp}. The line-of-sight average in one direction, $\vecn$, is then given by Eq.~\eqref{eq:I_av_los}.
For polarization, this is equivalent to performing a Taylor expansion of the spinor's modulus\footnote{Here we make the assumptions discussed in Sec.~\ref{sec:comments_pol}.} $\polnu=|\spinpol_\nu|$ with respect to the spectral parameters at each fixed polarization angle $\gamma$. However, we cannot use a perturbative approach to average over the polarization angles $\gamma$ since, in physical situations, one expects them to vary widely in a nontrivial way, such that the situation would quickly become mathematically inconsistent. This means that the line-of-sight average has to be generalized to include the polarization state in the parameter distribution function
%--------------------------------------
\begin{align}
\label{eq:P_av_los}
\langle \spinpol_\nu\rangle
&=\left< \polnu(A, \vec{p})\,\expf{2\i\gamma}\right>
\equiv\int \prob(\vec{p}, \gamma, \vecn) \,\hatpol_\nu(\vec{p})\,\expf{2\i\gamma} \,{\rm d}^N p\,{\rm d}\gamma.
\end{align}
%--------------------------------------
Here, we again used $\polnu(A, \vec{p})=A\,\hatpol_\nu(\vec{p})$, as for intensity.
In analogy to the intensity moment expansion, one then finds
%--------------------------------------
\begin{align}
     &\left<\spinpol_\nu(A,\vec{p}, \gamma)\right>
    = \hatpol_\nu(\barppolv)\left< A\,\expf{2\i\gamma}\right>
    + \sum_j^N \left< A\,\expf{2\i\gamma}(p_j-\barppol_j)\right>\, \partial_{\barppol_j}\hatpol_\nu(\barppolv)
    \nonumber \\
    &\qquad+ \frac{1}{2}\sum_{j,k}^N
    \left<A\,\expf{2\i\gamma} (p_j-\barppol_j)(p_k-\barppol_k)\right>\,
    \partial_{\barppol_j}\partial_{\barppol_k}\hatpol_\nu(\barppolv)
    + \dots\,, \nonumber
\end{align}
%--------------------------------------
which depends on the pivot $\barpv$, as we specify in Sect.~\ref{sec:pol_pivot}. Since $\left< A\,\expf{2\i\gamma}\right>$ can vanish, we cannot simply factor it out of the expressions. Instead, like for the intensity moments, we will again use $\bar A=\langle |A\,\expf{2\i\gamma}| \rangle\equiv \langle A \rangle$ as a way to normalize the distributions. This allows us to define the spin moments
%--------------------------------------
\begin{align}
\label{eq:spin-moments}
   \mathcal{W}_\alpha^{p_j\dots p_l} 
    =\frac{\left<A\,\expf{2\i\gamma} (p_j-\barppol_j)\dots(p_l-\barppol_l)\right>}{\barA}
    \equiv 
    \Omega_\alpha^{p_j\dots p_l} \expf{2\i \gamma_{\alpha}^{p_j\dots p_l}}
\end{align}
%--------------------------------------
very much like for intensity but with an extra spinor weight. The moments are now complex-valued and in the second step we expressed them in terms of the real numbers, $\Omega_\alpha^{p_j\dots p_l}$ and $\gamma_{\alpha}^{p_j\dots p_l}$. The latter defines average directions of polarization states associated with each of the moments of the SED. While considering the pair $(\Omega_\alpha,\gamma_\alpha)$ or the real and imaginary parts of $\mathcal{W}_\alpha$ give perfectly equivalent descriptions, one could be favored over the other for parameters estimation or physical interpretation. From a numerical perspective, considering the pair ($Q$,$U$) as the components of a single object instead of two independent intensities will add correlations between their moments, which is expected to improve the accuracy of the parameter inference. The final polarization moment expansion then takes the form:
%--------------------------------------
\begin{align}
\label{eq:momentgeneral}
    &\left<\spinpol_\nu(A,\vec{p}, \gamma)\right> 
    =  \mathcal{W}_0\,\polnu(\barA,\barppolv)
    + \sum_j \mathcal{W}_1^{p_j} \partial_{\barppol_j}\polnu(\barA,\barppolv)
    \nonumber \\  
    &\qquad\quad
    + \frac{1}{2}\sum_{j,k} \mathcal{W}_2^{p_j p_k} 
    \partial_{\barppol_j}\partial_{\barppol_k}\polnu(\barA,\barppolv)+\dots 
    \\ \nonumber
    &\qquad\qquad
    +\frac{1}{\alpha!} \sum_{j,\dots,l} \mathcal{W}_\alpha^{p_j\dots p_l}
    \partial_{\barppol_j}\dots\partial_{\barppol_l}\polnu(\barA,\barppolv)+ \dots,
\end{align}
%--------------------------------------
which in this form can be interpreted as the sum of multiple SEDs with well-defined polarization states. It is this sum of well-defined single polarization states (i.e., defined by the complex-valued moments) with varying SEDs (i.e., the derivative spectra) that leads to rotation of polarization planes. 

We comment that the number of parameters in Eq.~\eqref{eq:momentgeneral} depends on the moment order that is used in the modeling. For each moment, two degrees of freedom are added (i.e., the real and imaginary parts). In addition, one has to determine the spectral parameter pivot, $\barppolv$. However, the overall normalization $\bar{A}$ does not independently contribute, but was merely chosen to scale the moments. As such, it cannot be independently estimated, and only the values of $\bar{A}\,\mathcal{W}_\alpha$ actually matter.

\vspace{-3mm}
\subsubsection{Average polarization angle}
\label{sec:pol_angle}
%--------------------------------------
Since $\mathcal{W}_0=\Omega_0 \expf{2\i\gamma_{0}}=\left<A\,\expf{2\i\gamma}\right>/\barA$ can generally vanish, there is no longer a trivially defined average polarization angle. In particular when $\mathcal{W}_0\approx0$, the average polarization angle can be fully determined by the higher order terms in Eq.~\eqref{eq:momentgeneral} and also generally becomes frequency-dependent.

To illustrate this aspect, let us consider the simple example of two power law spectra with equal weights $A$ along the $\pm Q$ direction ($\gamma_1=0$ and $\gamma_2=\pi/2$). For these we have $\expf{2\i\gamma_1}=1$ and $\expf{2\i\gamma_2}=-1$ implying $\langle\spinpol_\nu\rangle=A(\nu/\nu_0)^{\beta_1}-A(\nu/\nu_0)^{\beta_2}$. For $\beta_1\neq \beta_2$, we find $\spinpol_\nu\neq 0$ unless $\nu\equiv \nu_0$, which is reflected by the fact that $\mathcal{W}_0=\left<A\,\expf{2\i\gamma}\right>/\barA\equiv (A-A)/[2A]=0$, implying that the leading order term in Eq.~\eqref{eq:momentgeneral} vanishes. Also, no matter what the frequency, in our example the polarization state will remain $Q_\nu$, with a change of sign at $\nu=\nu_0$ and hence flip of $0\leftrightarrow \pi/2$. In this situation, all higher order moments remain real and $\gamma_\nu$ is highly non perturbative (i.e., not differentiable) at $\nu=\nu_0$.

There must be a way to define a meaningful average polarization angle for each of the moment terms. Indeed, if we simply think of the average of $\gamma$ along the line of sight in terms of the distribution, $\prob(\vec{p}, \gamma, \vecn)$. This then results in
%--------------------------------------
\begin{align}
\gammab=\frac{\left<A \gamma\right>}{\left<A\right>}
=\frac{\int \prob(\vec{p}, \gamma, \vecn) \,\gamma \,{\rm d}^N p\,{\rm d}\gamma}{\int \prob(\vec{p}, \gamma, \vecn) \,{\rm d}^N p\,{\rm d}\gamma}
\end{align}
%---------------------------------------
as the average polarization angle. This angle can also be used as a pivot when expanding the polarization state: 
%--------------------------------------
\begin{align}
\label{eq:pol_angle_perturb}
\expf{2\i\gamma}&= \expf{2\i\gammab}\left[1+\sum_{k=1}^\infty \frac{(2\i)^k}{k!} \,(\gamma-\gammab)^k\right).
\end{align}
%--------------------------------------
Using this in Eq.~\eqref{eq:spin-moments}, have
%--------------------------------------
\begin{align}
\label{eq:spin-moments_II}
\mathcal{W}_\alpha^{p_j\dots p_l} 
%&=\expf{2\i\gammab}\,\frac{\left<A\,\expf{2\i(\gamma-\gammab)} (p_j-\barppol_j)\dots(p_l-\barppol_l)\right>}{\barA}
%\nonumber\\
&=\expf{2\i\gammab}\,\sum_{k=0}^\infty \frac{(2\i)^k}{k!} \frac{\left<A (\gamma-\gammab)^k (p_j-\barppol_j)\dots(p_l-\barppol_l)\right>}{\barA}.
\end{align}
%--------------------------------------
The first term in the sum (i.e., $k=0$), is the only non vanishing contribution if the distributions of $\gamma$ and $\vec{p}$ factorize (i.e., the two are uncorrelated variables), as we discuss in Sect.~\ref{sec:depolarization}. Adding term by term in the series of Eq.~\eqref{eq:spin-moments_II} allows us to include information from higher order correlations of $\gamma$ and $\vec{p}$. However, in terms of {\it distinguishable} parameters, only the total moments, $\mathcal{W}_\alpha^{p_j\dots p_l}$, can really be constrained.

\subsubsection{Definition of the pivot}
\label{sec:pol_pivot}
%--------------------------------------
How do we determine the spectral parameter pivot? In the intensity case, we simply demanded the first moments to vanish to fix the pivot. For polarization,  this naively yields the condition
%--------------------------------------
\begin{align}
\langle A \,\expf{2\i\gamma} \rangle \barppolv= \langle A \,\expf{2\i\gamma} \vec{p}\rangle.
\end{align}
%--------------------------------------
However, since $\langle A \,\expf{2\i\gamma} \rangle$ can vanish, in general this cannot be a meaningful choice. 

Above, we already defined $\barA=\langle |A\,\expf{2\i\gamma}| \rangle$. In a similar manner, we can introduce the SED pivots as
%--------------------------------------
\begin{equation}
\label{eq:pivot_pol}
    \barppolv
    =\frac{\left<|A\,\expf{2\i\gamma}| \vec{p}\right>}{\left<|A\,\expf{2\i\gamma}|\right>}\equiv \frac{\left<A \vec{p}\right>}{\left<A\right>}=\frac{\int \prob(\vec{p}, \gamma, \vecn) \,\vec{p} \,{\rm d}^N p\,{\rm d}\gamma}{\int \prob(\vec{p}, \gamma, \vecn) \,{\rm d}^N p\,{\rm d}\gamma},
\end{equation}
%--------------------------------------
which is equivalent to the definition for the intensity moments. Physically, this means that we disregard the geometrical properties of $\spinpol_\nu$ and simply treat its modulus as an intensity.
%directionality of the $Q_\nu$ and $U_\nu$, and simply co-add them as {\it positive} intensities when defining the average SED. 
For our power-law example in Sect.~\ref{sec:pol_angle}, this means $\bar{\beta}=(A_1\beta_1+A_2\beta_2)/(A_1+A_2)$, which is fully analogous to the result of a simple intensity moment expansion. As we shall see below, this choice is well motivated and leads to a well-behaved polarization moment formalism.

In the perturbative regime however, $\Omega_0=|\langle A\expf{2\i\gamma}\rangle| \gg 0$, one can safely choose the complex pivot
%-------------------
\begin{equation}
\label{eq:pivot_pol2}
    \barppolv
    =\frac{\left<A\,\expf{2\i\gamma} \vec{p}\right>}{\left<A\,\expf{2\i\gamma}\right>} \, \Rightarrow \, \Delta \bar{p_j}=\frac{\left<A\,\expf{2\i\gamma} (p_j-\bar{p}_j)\right>}{\left<A\,\expf{2\i\gamma}\right>}= \frac{\mathcal{W}^{p_j}_1}{\mathcal{W}_0}
\end{equation}
%-------------------
While the spectral parameters $\vec{p}$ are real quantities, correcting by a complex number might seem incoherent. However, as we will discuss with examples, doing so is deeply relevant. While the real part of $\barppolv$ can be interpreted as real correction of $\vec{p}$, its complex part gives rise to the first order frequency dependence of the polarization angle $\gamma_\nu$ and can add some spectral modulation to the polarized intensity. 

\subsubsection{Independent angle distribution and de-polarization}
\label{sec:depolarization}
%--------------------------------------
In the definition of the line-of-sight average and spin moments, Eq.~\eqref{eq:P_av_los} and \eqref{eq:spin-moments}, we kept the parameter distribution function general. The discussion is greatly simplified if the probability distributions for the spectral parameters and the polarization angles can be considered as independent. In this case, one has\footnote{In doing so, we can use the normalizations $\int \prob(\vec{p}, \vecn){\rm d}^N \,p= \bar{A}$ and $\int \prob(\gamma, \vecn){\rm d}\,\gamma=1$.} $\prob(\vec{p}, \gamma, \vecn)\approx \prob(\vec{p}, \vecn)\, \prob(\gamma, \vecn)$, such that
%--------------------------------------
\begin{align}
\label{eq:P_av_los_independent}
\langle \spinpol_\nu\rangle
&=\int \prob(\vec{p}, \vecn) \,\hatpol_\nu(\vec{p})\,{\rm d}^N p\,\int \prob(\gamma, \vecn) \,\expf{2\i\gamma} \,{\rm d}\gamma.
\end{align}
%--------------------------------------
As this expression shows, spectral mixing and polarization angle averaging become completely independent, such that {\it no} frequency-dependent polarization angle can be expected. However, when summing over different physical conditions along the line of sight, the probability distribution becomes
\begin{align}
    \prob(\vec{p}, \gamma, \vecn) &= \int A(s)\delta(\gamma-\gamma(s))\delta^{\rm N}(\vec{p}-\vec{p}(s)){\rm d} s \nonumber\\
    &\neq \int  \delta(\gamma-\gamma(s)){\rm d} s \int A(s) \delta^{\rm N}(\vec{p}-\vec{p}(s)){\rm d} s ,
\label{eq:prob_los}
\end{align}
introducing an unavoidable dependence between the angles and the spectral parameters. This dependence disappears if either the polarization angle or the spectral parameters are constant in the line of sight, highlighting that a variation of both $\gamma$ and $\vec{p}$ is required to have a spectral dependence of the polarization angle.

If the angle distribution is Gaussian with average angle $\gammab(\vecn)$ and width $\sigma_\gamma(\vecn)$, then one finds
%--------------------------------------
\begin{align}
\label{eq:av_pol_angle}
\int \prob(\gamma, \vecn) \,\expf{2\i\gamma} \,{\rm d}\gamma
&=\expf{2\i\gammab(\vecn)}\,\expf{-2\,\sigma_\gamma^2(\vecn)}.
\end{align}
%--------------------------------------
This expression highlights that the dispersion of the angles leads to damping of the net polarization amplitude and ultimately complete depolarization if the distribution becomes too wide. In this case, a general perturbative expansion in $\Delta \gamma=\gamma-\gammab$, [see e.g. Eq.~\eqref{eq:pol_angle_perturb}] is unlikely to converge, but, as stressed already, does not add any new insight anyways.

\section{Canonical SEDs}
\label{sec:applications}
%--------------------------------------
In this section, we illustrate the spin-moment framework on some detailed analytical and numerical examples relevant to astrophysical applications. We consider discrete sums of polarized SEDs along a given line of sight, often focusing on very few contributions. For the moment formalism, this can lead to non-perturbative cases, since in the limit of many emitters, the moments are expected to become more Gaussian due to the central limit theorem. Still, in most cases only a few moments are required to capture the dominant effects.

To highlight the performance of the moment formalism, we treat the sum of SEDs with noise as data and then use the moment representations to finite order as model. We perform a parameter estimation by means of curve fitting with $\chi^2$ minimization in complex-variables using the {\tt LMFIT} python library \citep{Lmfit}. Hereafter, the model of linear polarization given by the spin-moment expansion is $\spinpol_\nu^{\rm M}$ and the simulated data signal is noted $\spinpol_\nu^{\rm S}$. We add Gaussian noise $\mathcal{N}_\nu$ to the simulation, with zero mean and standard deviation $\sigma = \sigma_Q + \i \sigma_U$. The values of $\sigma$ is chosen such that the signal to noise ratios $Q_\nu^{\rm S}/Q_\nu^{\rm M}$ and $U_\nu^{\rm S}/U_\nu^{\rm M}$ are constants over the whole frequency range (chosen arbitrarily to be $1\times 10^{-5}$). \footnote{The error bars used in all the figures are respectively given by $\sigma_Q^2$, $\sigma_U^2$, $\sigma_P^2=(Q^2\sigma_Q^2+U^2\sigma_U^2)/P^2$ and $\sigma^2_\gamma=0.5(U^2\sigma_Q^2+Q^2\sigma_U^2)/P^4$.}

The $\chi^2$ to minimize is given by $\chi^2 = \frac{1}{2}|\spinpol_\nu^{\rm M} -\spinpol_\nu^{\rm S}|^2/|\sigma|^2$. The signal is considered over a frequency range going from 1\,GHz to $\nu_{\rm max}$ in intervals of 1\,GHz. The choice of $\nu_{\rm max}$ will depend on the example considered. We introduce the shorthand notation '$O(\alpha)$' to refer to the fit of the spin-moment expansion including all the terms up to order $\alpha$. '$O(0)$' is the leading order/canonical SED.
Two distinct routines are developed, fitting either the pair $({\rm Re}(\mathcal{W}_\alpha^{\vec{p}}),{\rm Im}(\mathcal{W}_\alpha^{\vec{p}}))$ or the pair $(\Omega_\alpha^{\vec{p}},\gamma_\alpha^{\vec{p}})$. In all the examples considered, both lead to identical results and we leave a further comparison between the two implementations for future work.
We are interested in the SED distortions and their behavior in the complex plane, which are only driven by the relative contributions of the different emission points. As such, we use natural units of $\Iunit$, for all the SEDs. A more detailed discussion on the relevance of weights, normalization and change of units can be found in Appendices~\ref{sec:normalization} and \ref{sec:weights}.

\subsection{General discrete sums of canonical SEDs}
\label{sec:manySEDs}
%--------------------------------------
For a discrete sum of $M$ SEDs along a line of sight one can trivially write the distribution function as
%--------------------------------------
\begin{align}
\prob(\vec{p}, \gamma, \vecn) 
= \sum_k^M A_k\, \delta(\gamma-\gamma_k)\,\delta^N(\vec{p}-\vec{p}_k),
\end{align}
%--------------------------------------
where $\delta(x-x_0)$ denotes Dirac's distribution and the sum extends over the discrete emission points along the line of sight with SED vectors $\vec{p}_k$ and polarization angles $\gamma_k$. Inserting this into the definitions of the moments and pivots given in the previous section we trivially find the exact average
%--------------------------------------
\begin{align}
\langle\spinpol_\nu\rangle &= 
\sum_k^M A_k \int \delta(\gamma-\gamma_k)\,\delta^N(\vec{p}-\vec{p}_k) \,\hatpol_\nu(\vec{p})\,\expf{2\i\gamma} \,{\rm d}^N p\,{\rm d}\gamma
\nonumber\\
&=\sum_k^M A_k \expf{2 \i \gamma_k} \hatpol_\nu(\vec{p}_k).
\end{align}
%--------------------------------------
Using the polarization moment expansion, we automatically have the normalization, pivot and complex-valued moments as
%--------------------------------------
\begin{subequations}
\vspace{-4mm}
\begin{align}
    \barA&= 
    \sum_k^M A_k \int \delta(\gamma-\gamma_j)\,\delta^N(\vec{p}-\vec{p}_j)\,{\rm d}^N p\,{\rm d}\gamma= \sum_k^M A_k 
    \\
\barppolv&=\sum_k^M (A_k/\barA)\,\vec{p}_k, 
\quad
\gammab=\sum_k^M (A_k/\barA)\,\gamma_k
    \\
\mathcal{W}_0
    &=\sum_k^M (A_k/\barA)\, \expf{2\i\gamma_k}
\\
\mathcal{W}_\alpha^{p_j\dots p_l} 
    &=\sum_k^M (A_k/\barA)\,\expf{2\i\gamma_k} (p_{k,j}-\barppol_{k,j})\dots(p_{k,l}-\barppol_{k,l}),
\end{align}
\end{subequations}
%--------------------------------------
where the ratios $A_k/\barA$ determine the probabilities to find $\vec{p}_k$ and $\gamma_k$. These expressions can then be inserted into Eq.~\eqref{eq:momentgeneral} to obtain the polarization moment expansion. The derivatives of the spectra have to be computed individually, but generally the moment expansion is expected to converge with only a few terms.

\subsection{Power laws \label{sec:power-laws}}
%--------------------------------------
As a first example of astrophysical relevance, we consider the simple case of power-law SEDs:
%--------------------------------------
\begin{equation}
    \hatpol_\nu^{\,\rm PL}(\beta) = \fracnu^\beta.
\end{equation}
%--------------------------------------
The polarization state can then be characterized by $\spinpol^{\rm PL}_\nu\approx A\expf{2\i\gamma}\hatpol_\nu^{\,\rm PL} $. The only spectral parameter relevant for the moment expansion is the spectral index $\vec{p}=(\beta)$, normalized at a reference frequency $\nu_0$.
This SED plays a crucial role in the foreground modeling of synchrotron on large scales \citep{PlanckCompoSep}. In the following numerical applications, we choose $\nu_{\rm max}=150$\,GHz below which the synchrotron emission is dominant and $\nu_0=23$\,GHz as the \wmap{} frequency band \citep{WMAPfg}. 
%\jenscomment{fix line spill.}
%
Using $\partial_\beta^k \hatpol^{\rm PL}_\nu=\partial_\beta^k (\nu/\nu_0)^\beta=(\nu/\nu_0)^\beta \ln(\nu/\nu_0)^k$, the spin-moment expansion in  Eq.~\eqref{eq:momentgeneral} can then be expressed as
%--------------------------------------
\begin{align}
\label{eq:powerlawexp}
\langle\spinpol^{\,\rm PL}_\nu\rangle 
&= \polnu^{\,\rm PL}(\barA, \bar{\beta}) \times 
\Bigg\{\mathcal{W}_0+ \mathcal{W}^{\beta}_1 \lnnu \nonumber\\ &\qquad\qquad+\frac{\mathcal{W}^{\beta^2}_2}{2}\lnnu^2
+\frac{\mathcal{W}^{\beta^3}_3}{6}\lnnu^3+  \cdots \Bigg\}.
\end{align}
%-------------------------------------
The choice of the reference frequency $\nu_0$, around which to make the expansion, can have an impact on the convergence rate of model, but otherwise leaves the moment expansion unchanged. One choice is to pick a local extremum where the SED changes shape: $\partial_{\nu}\polnu^{\rm S} =0$ or $\partial_{\nu}\gamma^{\rm S}_\nu =0$ depending on the distortion type. In front of real data the choice has to be made also from instrumental considerations.
%--------------------------------------
As an example, we now consider the superposition of two power laws in more detail.

\subsubsection{Hands-on example: two power laws}
%--------------------------------------
Consider two power laws ($M=2$) with different spectral indices $(\beta_1,\beta_2)$ and polarization angle $(\gamma_1,\gamma_2)$ along the same line of sight. The exact solution then reads
%--------------------------------------
\begin{align}
\spinpol_\nu = A_1 \fracnu^{\beta_1} \expf{2\i \gamma_1} + A_2 \fracnu^{\beta_2} \expf{2\i \gamma_2}. 
\end{align}
%--------------------------------------
Carrying out the intensity moment expansion of the two individual power laws with respect to their spectral indices, we obtain
%--------------------------------------
\begin{subequations}
\label{eq:PL_expansion}
\vspace{-4mm}
\begin{align}
\barA&=A_1+A_2
\\
\bar{\beta}&=\frac{A_1}{\barA}\,\beta_1+\frac{A_2}{\barA}\,\beta_2, 
\quad 
\gammab=\frac{A_1}{\barA}\,\gamma_1+\frac{A_2}{\barA}\,\gamma_2
\\
\mathcal{W}_0&=\frac{A_1}{\barA} \expf{2\i\gamma_1}+\frac{A_2}{\barA} \expf{2\i\gamma_2}
\\
\mathcal{W}_\alpha^{\beta^\alpha} 
    &=\frac{A_1}{\barA}\,\expf{2\i\gamma_1} (\beta_1-\bar{\beta})^\alpha
     +\frac{A_2}{\barA}\,\expf{2\i\gamma_2} (\beta_2-\bar{\beta})^\alpha.
\end{align}
\end{subequations}
%--------------------------------------
These expressions can be trivially extended to $M$ power laws after extending the sums to $M$ parameters $A_k, \gamma_k$ and $\beta_k$ to find the values of the pivot and spin moments (see Sect.~\ref{sec:manySEDs}). However, for illustrations the two power-law case is more intuitive.

If $\gamma_1=\gamma_2=\bar{\gamma}$ and $\beta_1\neq \beta_2$, we naturally find that all spin moments are aligned in the same directions of the complex-plane and hence no change in the polarization direction can occur as a function of frequency. In this case, $\langle \spinpol^{\rm PL}_\nu \rangle=\expf{2\i \gamma}\langle \polnu^{\rm PL}(A,\beta) \rangle$, trivially describing the effect of spectral mixing only. If on the other hand $\beta_1=\beta_2=\bar{\beta}$ and $\gamma_1\neq\gamma_2$, we naturally have $\mathcal{W}_\alpha^{\beta^\alpha}=0$ and everything is described by $\mathcal{W}_0$ with a fixed SED. 
%This is the case of pure polarization angle mixing.
To obtain nontrivial consequences of polarized mixing, both $\gamma_k$ and $\beta_k$ need to vary.

%----------------------------------
\begin{figure}[ht]
    \centering
    \includegraphics[width=0.92\columnwidth]{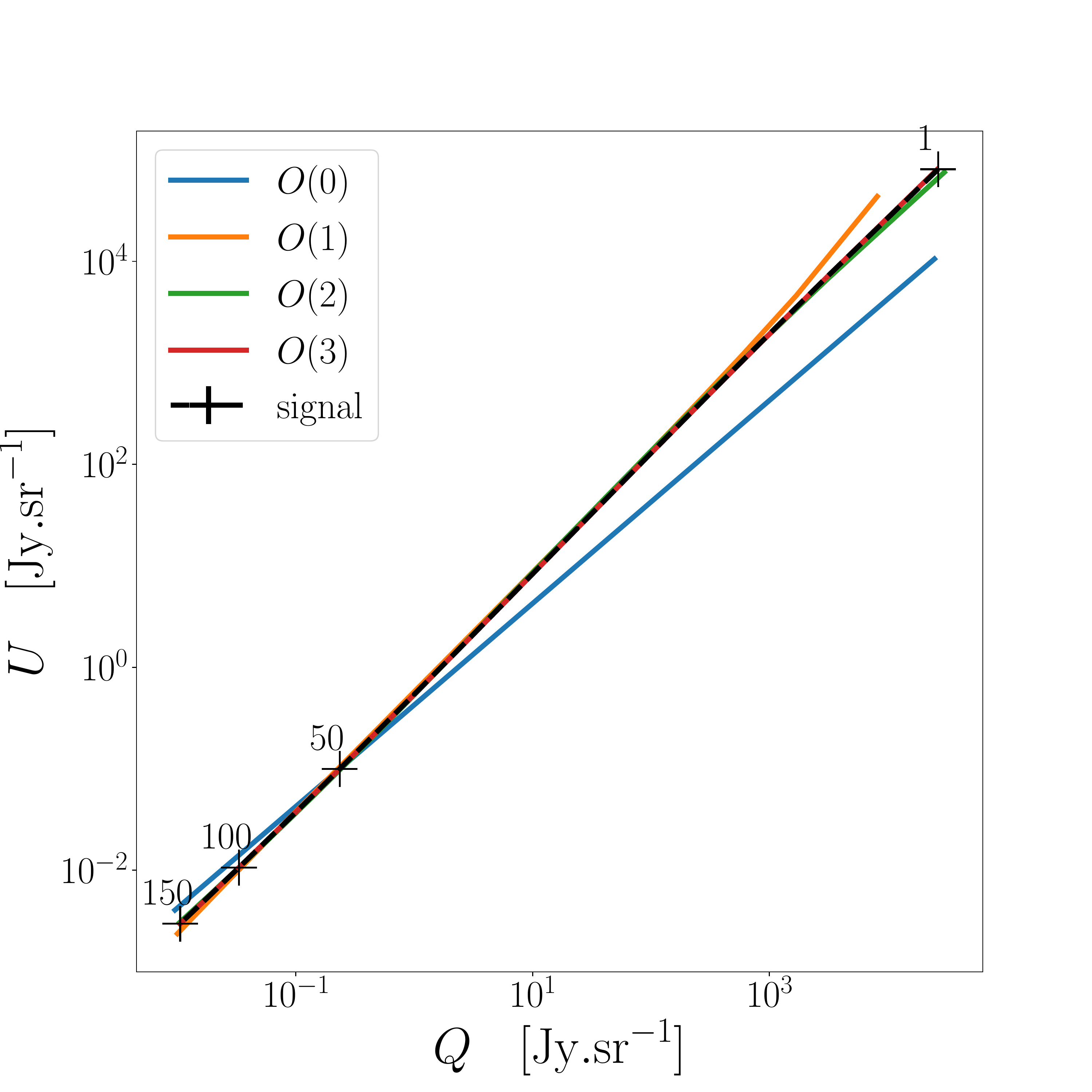}
    \includegraphics[width=0.92\columnwidth]{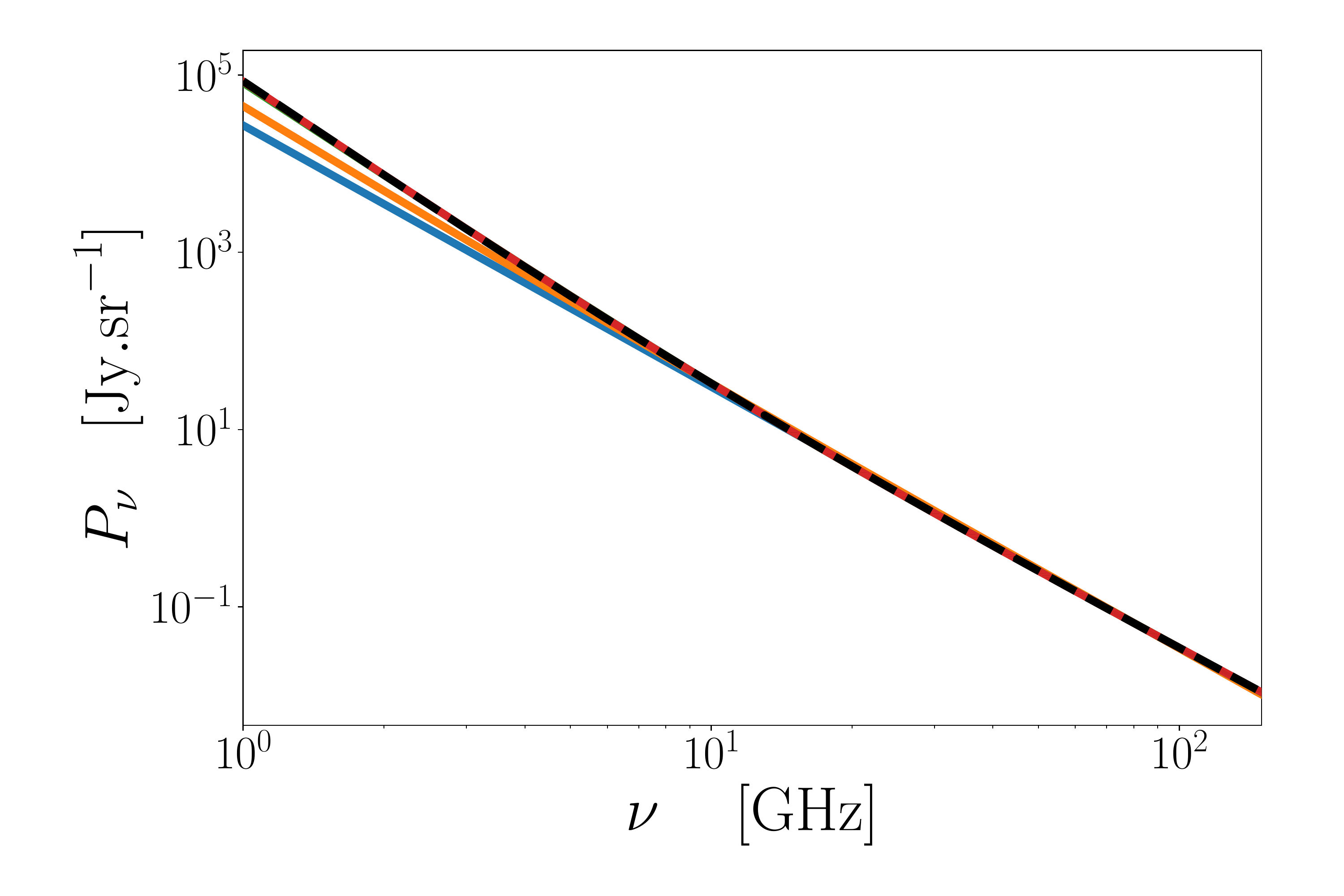}
    \includegraphics[width=0.92\columnwidth]{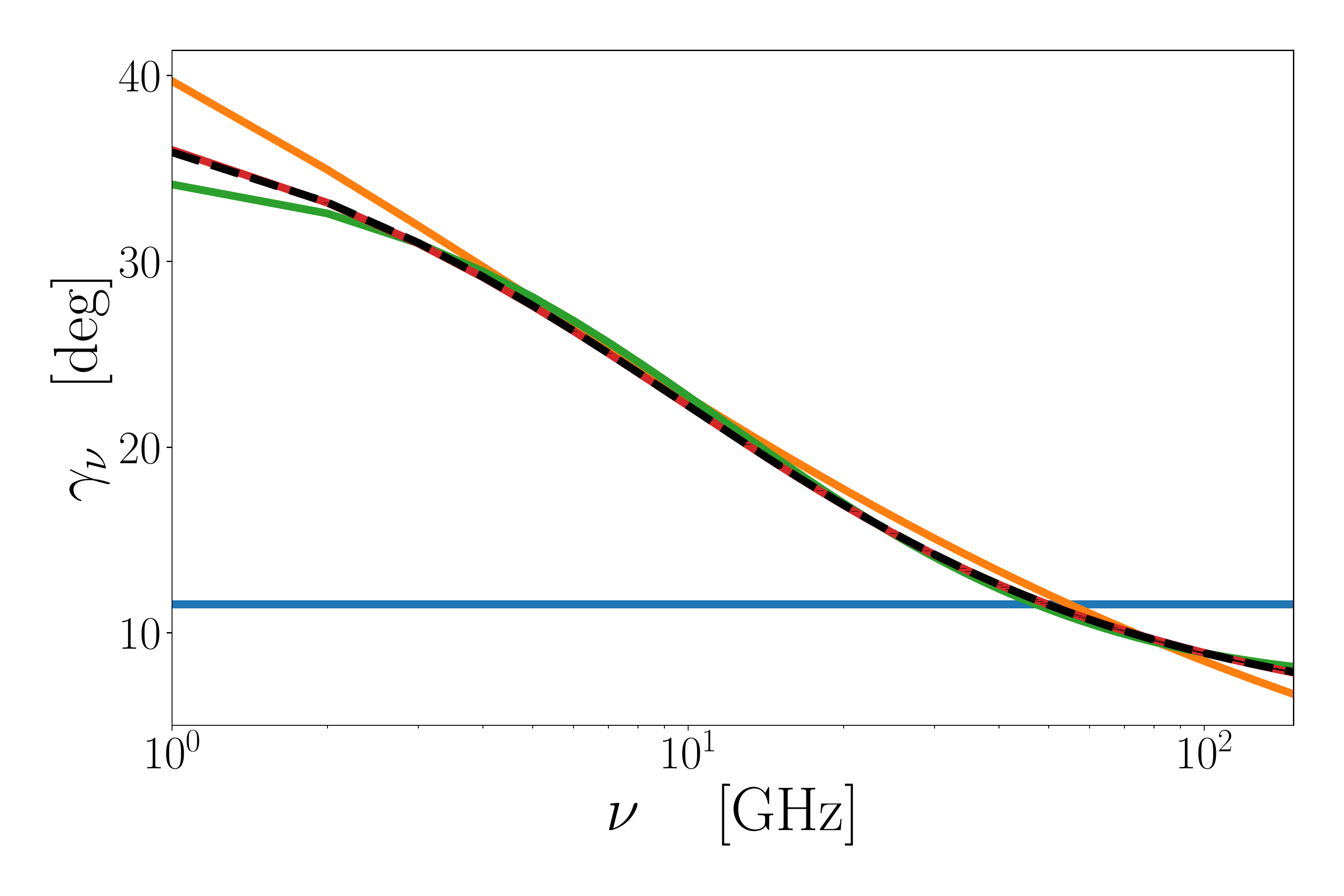}
    \caption{\footnotesize 
    Illustration of the spinor $\spinpol_\nu$ in the complex plane $(Q,U)$ for a sum of two power laws ({\it upper panel}). Black crosses mark the steps of 50 GHz on the signal. The values of frequencies are indicated above the crosses in GHz. The corresponding polarized intensity $\polnu$ ({\it central panel}) and polarization angle $\gamma_\nu$ ({\it lower panel}).
    % with parameters $A_1=1.2\Iunit$, $A_2= 2\Iunit$, $\beta_1=2$, $\beta_2=0.5$,
    % $2\gamma_1 = 72^{\circ}$,
    % $2\gamma_2 = -90^{\circ}$.
    The exact result with associated (invisible) error bars (black) is compared to the best-fit moment representation at various orders.
    %: leading order $O(0)$ (blue), first order $O(1)$ (orange), second order $O(2)$ (green) and third order $O(3)$ (red). 
    }
    \label{fig:2PL-P-gamma}
\end{figure}
%----------------------------------

%----------------------------------
As an illustration, let us consider the highly non-perturbative example, with synchrotron-like behavior: $A_1=2\Iunit$, $A_2= 1\Iunit$, $\beta_1=-2.8$, $\beta_2=-3.6$, $2\gamma_1=10^{\circ}$ and $2\gamma_2= 80^{\circ}$, implying $\bar{A}=3\Iunit$,  $\bar{\beta}\approx -3.06$ and $\gammab\approx 16.6^{\circ}$. 
In Fig.~\ref{fig:2PL-P-gamma}, the modulus $\polnu$ and the argument $\gamma_\nu$ of the signal together with the recovered spinor representation are displayed for various orders of the expansion going from $O(0)$ to $O(3)$. 
%
% One can see that in the $(Q,U)$ plane a strong turn is taken by the spinor at low frequencies. The leading order does not capture this behaviour since it represents $\gamma_\nu\approx 0.5$.
%, crossing the signal at $\nu=\nu_0$.
%
For such strong deviations of spectral parameters, one cannot expect to find $\bar{\beta}\approx -3.06$ for the expansion at leading order, and it has to be treated as a free parameter of the model. This is particularly important when only a few moment terms are included.
The expansion at higher orders then allows us to gradually recover the nontrivial polarization signal over the frequency range. In all cases but $O(0)$, the best fit values for $\bar{A}$, $\bar{\beta}$ and $\mathcal{W}^\beta_\alpha$ are all compatible within one standard deviation with those given by Eq.~\eqref{eq:PL_expansion}.
While, by definition, the leading order cannot encompass any rotation of the spinor with frequency, we can see that the moment expansion allows us to correctly model the frequency dependence of $\gamma_\nu$. 

\subsubsection{Extreme cases and perturbative regime}
%----------------------------------
To gain further insight, 
let us just consider the first order terms of the expansion:
%--------------------------------------
\begin{align}
\label{eq:powerlawexp2}
\langle\spinpol^{\,\rm PL}_\nu\rangle 
&\approx \hatpol_\nu^{\,\rm PL}(\barA, \bar{\beta}) \times 
\Bigg\{\mathcal{W}_0+ \mathcal{W}^{\beta}_1 \lnnu \Bigg\}.
\end{align}
%--------------------------------------
If $\mathcal{W}_0\approx 0$, we indeed find the situation where we have a polarization angle fully determined by $\mathcal{W}^{\beta}_1$, with a sign-flip at $\nu=\nu_0$. 
The polarization SED is then determined by the first $\beta$ derivative of the power law, and polarization rotation would stem from higher order moments which are not included here. In this situation, we are dealing with two dominant (and near degenerate) contributions to the polarization state that are rotated by $90^\circ$ to each other (e.g., $+Q_\nu$ and $-Q_\nu$). The moment expansion then describes how much these two power-law terms differ.

%--------------------------------------
\begin{figure}[ht]
    \centering
    \includegraphics[width=0.92\columnwidth]{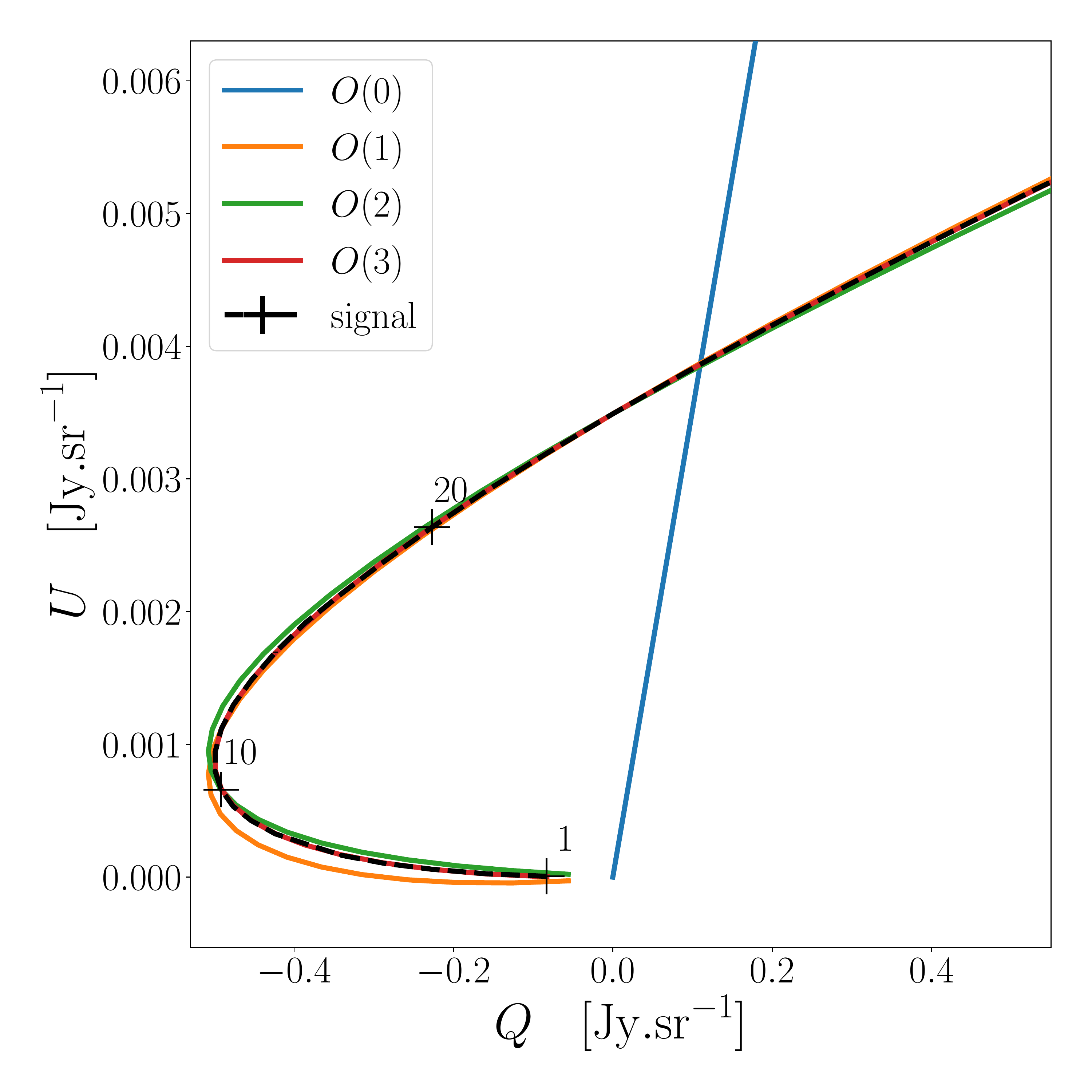}
    \includegraphics[width=0.92\columnwidth]{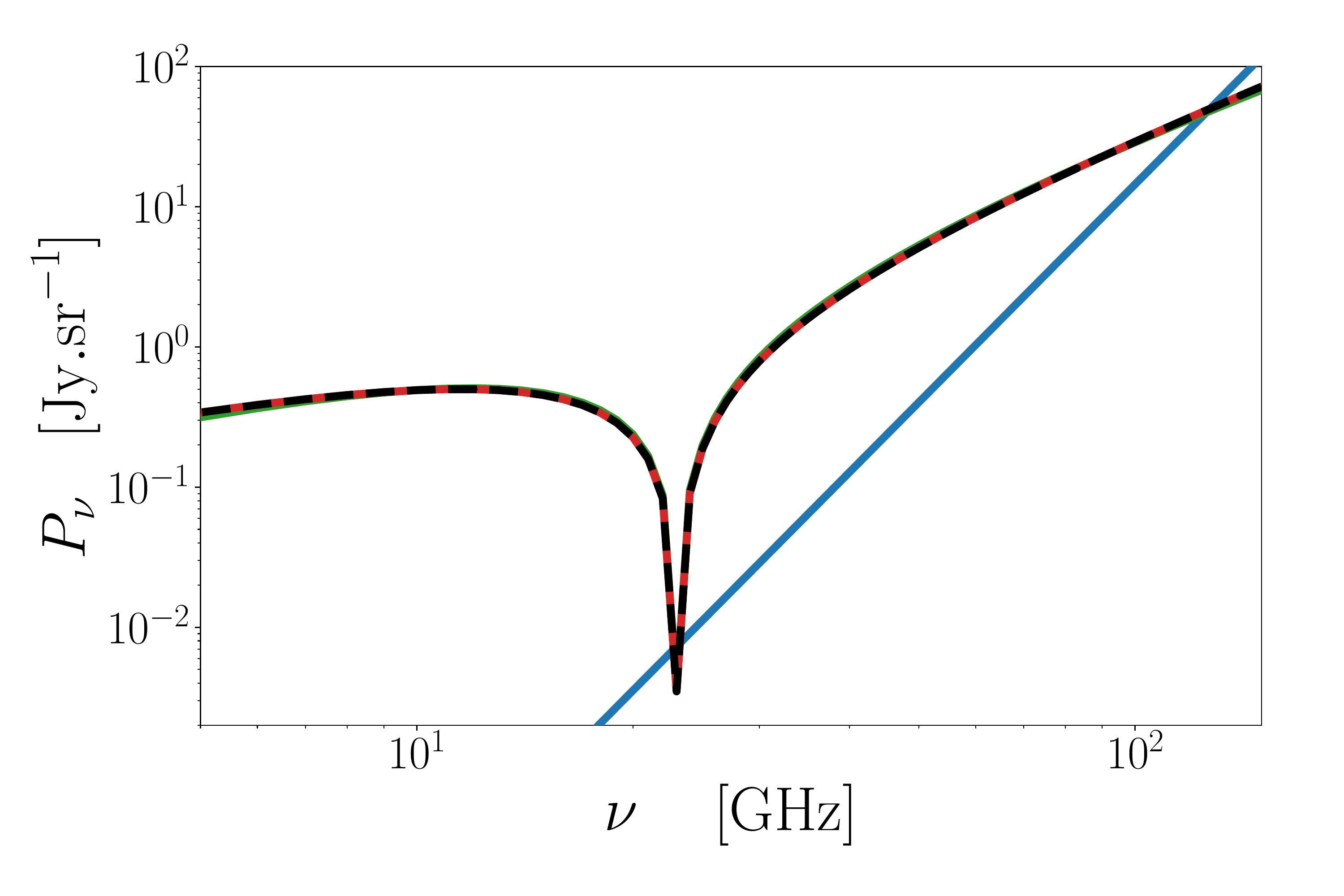}
    \includegraphics[width=0.92\columnwidth]{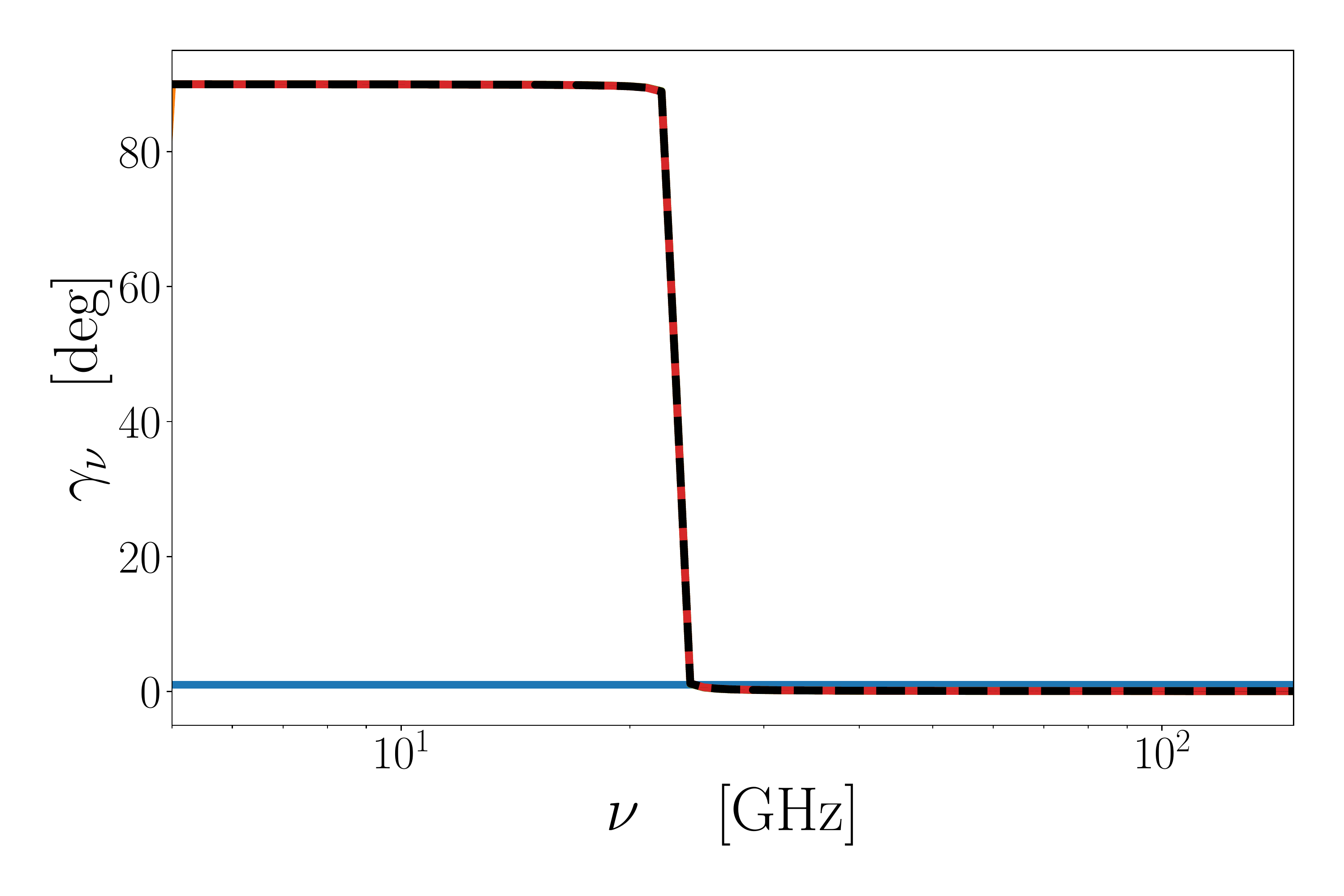}
    \caption{ Illustration of the spinor $\spinpol_\nu$ in the complex plane $(Q,U)$ for a sum of two almost anti aligned power laws, with a close-up view around $\nu=\nu_0$ ({\it Upper panel}). Black crosses mark the steps of 10 GHz on the signal.The values of frequencies are indicated above the crosses in GHz. The corresponding polarized intensity $\polnu$ ({\it Central panel}) and polarization angle $\gamma_\nu$ ({\it Lower panel}).
    %
    % with parameters $A_1=A_2=2$, $\beta_1=1$, $\beta_2=2$, $2\gamma_1 = 180^{\circ}$ and $0.1^{\circ}\simeq 0^{\circ}$. 
    %
    We choose logarithmic representations to emphasize the focus on the $\nu\sim\nu_0$ point. 
    %
    % The exact result with associated (invisible) error-bars (black) compared to the best-fitting moment representation at various orders: leading order (blue), 1. order (orange), 2. order (green) and 3. order (red).}
    }
    \label{fig:2GB-P-PLflip}
\end{figure}
%--------------------------------------

This situation is illustrated in Fig.~\ref{fig:2GB-P-PLflip} by a sum of two power laws with parameters $A_1=A_2=2\Iunit$, $\beta_1=1$, $\beta_2=2$, $2\gamma_1 = 180^{\circ}$ and $2\gamma_2=0.1^{\circ}$. One can see that both $\gamma_\nu$ and $\polnu$ do not behave as smooth functions at the breaking point $\nu \simeq \nu_0$, where one power law abruptly takes over the other one. Here, $\gamma_\nu$ changes very rapidly from $\gamma_1$ to $\gamma_2$. This is, however, not a problem for the spin-moment expansion, which allows us to recover $\spinpol_\nu$ correctly. As mentioned above, one recovers a very small best-fit value for $\mathcal{W}_0$ and $\mathcal{W}^\beta_1 \neq 0$ is the dominant term of the expansion, and polarization rotation mainly stems from the second and higher order moments.

If on the other hand we are in the perturbative regime where $|\mathcal{W}_0|\gg |\mathcal{W}^{\beta}_1|>0$, we can correct for the first order term of the expansion as in intensity, interpreting it as a correction to the spectral parameters with this correction now being complex-valued. let us split its real and imaginary parts as
%--------------------------------------
\begin{align}
\label{eq:Deltabeta}
\Delta\bar{\beta}=\frac{\mathcal{W}^{\beta}_1}{\mathcal{W}_0}=a_{\Delta\bar{\beta}}+\i b_{\Delta\bar{\beta}}.
\end{align}
%--------------------------------------
With \footnote{In general $\Omega_0$ is expected to depart from unity.} $\mathcal{W}_0=\Omega_0\,\expf{2\i \gamma_0}$, we can then write the expansion as
%--------------------------------------
\begin{align}
\langle\spinpol^{\,\rm PL}_\nu\rangle 
&\approx \mathcal{W}_0\,\hatpol_\nu^{\,\rm PL}(\barA, \bar{\beta}) \times 
\Bigg\{1+ \Delta\bar{\beta} \lnnu \Bigg\}
\nonumber\\
&\approx 
\barA\,\mathcal{W}_0\,
\fracnu^{\bar{\beta}+a_{\Delta\bar{\beta}}+\i\,b_{\Delta\bar{\beta}}}\nonumber\\
    &\equiv \barA\,\Omega_0\,
    \exp\left(2\i\left[\gamma_0+\frac{b_{\Delta\bar{\beta}}}{2}\,\ln\left(\frac{\nu}{\nu_0}\right)\right]\right)\,
    \fracnu^{\bar{\beta}+a_{\Delta\bar{\beta}}}.
\label{eq:PLpivotcorr}
\end{align}
%--------------------------------------
In the perturbative regime, we thus find a polarization SED that is again a power law with a spectral index and frequency-dependent polarization angle given by
%--------------------------------------
\begin{subequations}
\label{eq:perturbativepivotPL}
\begin{align}
\bar{\beta}^{\rm PL}&\approx \bar{\beta}+{\rm Re}\left(\frac{\mathcal{W}^{\beta}_1}{\mathcal{W}_0}\right)
\\
\gamma^{\rm PL}_\nu&\approx \gamma_0+\frac{1}{2}\,{\rm Im}\left(\frac{\mathcal{W}^{\beta}_1}{\mathcal{W}_0}\right)\,\ln\left(\frac{\nu}{\nu_0}\right).
\end{align}
\end{subequations}
%--------------------------------------
This result clearly shows that even at the lowest order, the superposition of linearly polarized power-law SEDs generally leads to a frequency-dependent rotation of the polarization angle. The rotation is purely driven by the imaginary part of the pivot value in the power-law index. At lowest order, no spectral curvature is added, even if the spectral index departs from that of the simple intensity superposition, $\bar{\beta}$, by $a_{\Delta\bar{\beta}}$. 
At higher order, the complex-valued moments given in Eq.~\eqref{eq:PL_expansion} lead to additional spectral complexity, which in most relevant situations can be captured in a perturbative manner.

%---------------------------------
Just as for intensity, in the perturbative regime one can cancel $\mathcal{W}^\beta_1$ and correct the leading order according to Eq.~\eqref{eq:PLpivotcorr}. Proceeding iteratively on numerical examples, one can witness the quick convergence of the first order moment toward zero. Doing so allows us to find the right pivot for the expansion while still keeping the pivot $\bar{\beta}$ fixed. As such, one breaks unwanted degeneracies and recovers a physically relevant value for the spectral parameters with a minimal dispersion.

\subsection{Blackbodies}
%--------------------------------------
As the second example, we briefly consider the superposition of blackbody spectra, a case that is directly relevant to primordial CMB polarization. The only free parameter is the blackbody temperature, and a temperature difference in two orthogonal directions is required to obtain a net polarization. This is an example where the origin of the polarization is due to the spectral parameters {\it only}. The SED for polarized light, at leading order in the temperature perturbation around the average, is then given by the first temperature derivative of a blackbody:
%--------------------------------------
\begin{align}
 B_\nu(\bar{T})&=\frac{2 h }{c^2} \frac{\nu^3}{\expf{h\nu/k\bar{T}} - 1},
 \quad
 %\nonumber\\
 G_\nu(\bar{T})
 &=\frac{\partial B_\nu(\bar{T})}{\partial \ln \bar{T}}
 =\frac{2 h \nu^3}{c^2} \frac{x\,\expf{x}}{\left(\expf{x} - 1\right)^2},
\nonumber\\
\spinpol^{\,\rm BB}_\nu&\approx G_\nu(\bar{T})\left[\Theta_Q+\i\Theta_U\right]
\end{align}
%--------------------------------------
with $x=h\nu/k\bar{T}$ and the usual natural constants. We also introduced the two temperature perturbations $\Theta_Q=\Delta T_Q/\bar{T}$ and $\Theta_U=\Delta T_U/\bar{T}$, which are respectively defined for a coordinate system that is rotated by $45^\circ$.
At fixed $\Theta_Q$ and $\Theta_U$, this means that (at lowest order in the temperature fluctuations) no spectral mixing happens, and hence $\gamma$ remains frequency-independent.

However, if we include terms at second order in $\Theta$, one finds an additional frequency dependence that is characterized by a $y$-type distortion \citep[e.g., see Appendix~A of][]{Chluba2015tensors}:
%--------------------------------------
\begin{align}
Y_\nu(\bar{T})&
=\frac{2 h \nu^3}{c^2} \frac{x\,\expf{x}}{\left(\expf{x} - 1\right)^2}\,\left[x\frac{\expf{x} + 1}{\expf{x} - 1}-4\right],
\nonumber\\
\spinpol^{\,\rm BB}_\nu&= G_\nu(\bar{T})\left[\Theta_Q+2\Theta_I\Theta_Q
+\i\left(\Theta_U+2\Theta_I\Theta_U\right)\right]
\nonumber\\[1mm]
&\qquad\qquad+Y_\nu(\bar{T})\left[\Theta_I\Theta_Q
+\i\Theta_I\Theta_U\right].
\end{align}
%--------------------------------------
Here, we introduced the total intensity temperature perturbation $\Theta_I=\Delta T_I/\bar{T}$, which generally includes both polarized and unpolarized contributions. Depending on the ratio of $\Theta_Q$ to $\Theta_U$, this will cause a small frequency-dependent rotation of the polarization planes. However, since this effect is at second order in the (small) CMB temperature differences, we leave a more detailed discussion to future work.

\vspace{-1mm}
\subsection{Gray-body spectra}
%--------------------------------------
In contrast to the blackbody, a gray-body (GB) spectrum also has a free normalization, caused by imperfect reflectivity of the material. As discussed in Sec.~\ref{sec:comments_pol}, we only allow variations of $A$ to create polarization. The fundamental SED is thus given by $\hatpol^{\rm GB}_\nu \equiv B_\nu(T)$, such that the single polarization state can be characterized by $\spinpol^{\rm GB}_\nu\approx A\,\expf{2\i\gamma} B_\nu(T)$.

We can then consider general GB superpositions. Following \citet{Chluba}, we shall use $\betaGB=1/T$ as the spectral parameter. The SED derivatives then have a closed form using Eulerian numbers \citep{Chluba2012SZpack}, with the first few terms given by \citep[see Eq.~(38) of][]{Chluba}:
%--------------------------------------
\begin{subequations}
\begin{align}
\betaGB \partial_{\betaGB} \hatpol^{\rm GB}_\nu&=-\hatpol^{\rm GB}_\nu\,\frac{x\expf{x}}{(\expf{x}-1)}
\\
\betaGB^2 \partial^2_{\betaGB} \hatpol^{\rm GB}_\nu&=+\hatpol^{\rm GB}_\nu\,\frac{x\expf{x}}{(\expf{x}-1)}\,x\coth(x/2)
\\
\betaGB^3 \partial^3_{\betaGB} \hatpol^{\rm GB}_\nu&=-\hatpol^{\rm GB}_\nu\,\frac{x\expf{x}}{(\expf{x}-1)}\,x^2\frac{\cosh(x)+2}{\cosh(x)-1}
\\
\betaGB^4 \partial^4_{\betaGB} \hatpol^{\rm GB}_\nu&=+\hatpol^{\rm GB}_\nu\,\frac{x\expf{x}}{(\expf{x}-1)}\,\frac{x^3}{2}\frac{\cosh(x)+5}{\sinh^2(x/2)}\,\coth(x/2).
\end{align}
\end{subequations}
%--------------------------------------
with the frequency variable $x=\frac{h\nu}{k} \,\betaGBb\equiv \frac{h\nu}{k \bar{T}}$.

The final GB moment expansion then takes the form:
%--------------------------------------
\begin{subequations}
\begin{align}
\langle\spinpol^{\rm GB}_\nu\rangle 
&= \polnu^{\rm GB}(\bar{A},\bar{T})\times \bigg\{\mathcal{W}_0 
+ \mathcal{W}^{\betaGB}_1\,Y^{\rm GB}_1(x) + \frac{1}{2}\mathcal{W}^{\betaGB^2}_2\,Y^{\rm GB}_2(x)
\nonumber \\ 
&\qquad\qquad\qquad\qquad+ \frac{1}{6}\mathcal{W}^{\betaGB^3}_3 \,Y^{\rm GB}_3(x)+ \cdots \bigg\}
\\
\bar{A}&=\langle A\rangle, \quad
\mathcal{W}_0 = \frac{\langle A\,\expf{2\i \gamma}\rangle}{\bar{A}}, \quad
\betaGBb \equiv\frac{1}{\bar{T}}=\frac{\langle A\,\betaGB\rangle}{\bar{A}}
\\
\mathcal{W}^{\betaGB^\alpha}_\alpha&=
\frac{\langle A\,\expf{2\i \gamma}\,(\betaGB-\betaGBb)^\alpha\rangle}{\bar{A}}
, \quad Y^{\rm GB}_k(x)=\frac{1}{\hatpol^{\rm GB}_\nu}\,\frac{\partial^k\hatpol^{\rm GB}_\nu}{\partial \betaGBb^k}.
\end{align}
\end{subequations}
%--------------------------------------
The functions $Y^{\rm GB}_k(x)$ will also be relevant to the discussion of modified blackbody spectra in Sect.~\ref{sec:mBB_pol}.

%--------------------------------------
\begin{figure}[t!]
    \centering
    \includegraphics[width=0.92\columnwidth]{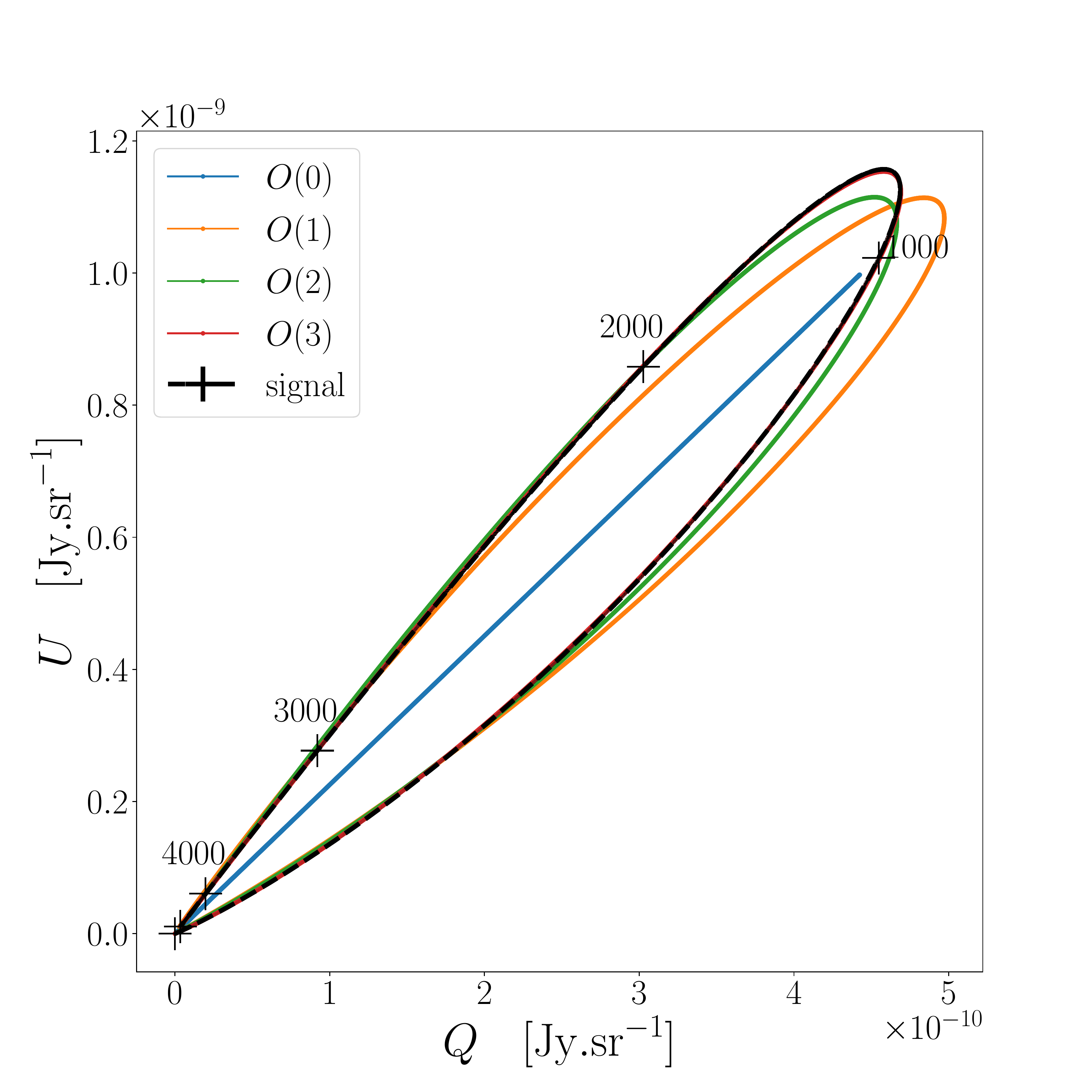}
    \\[1mm]
    \includegraphics[width=0.92\columnwidth]{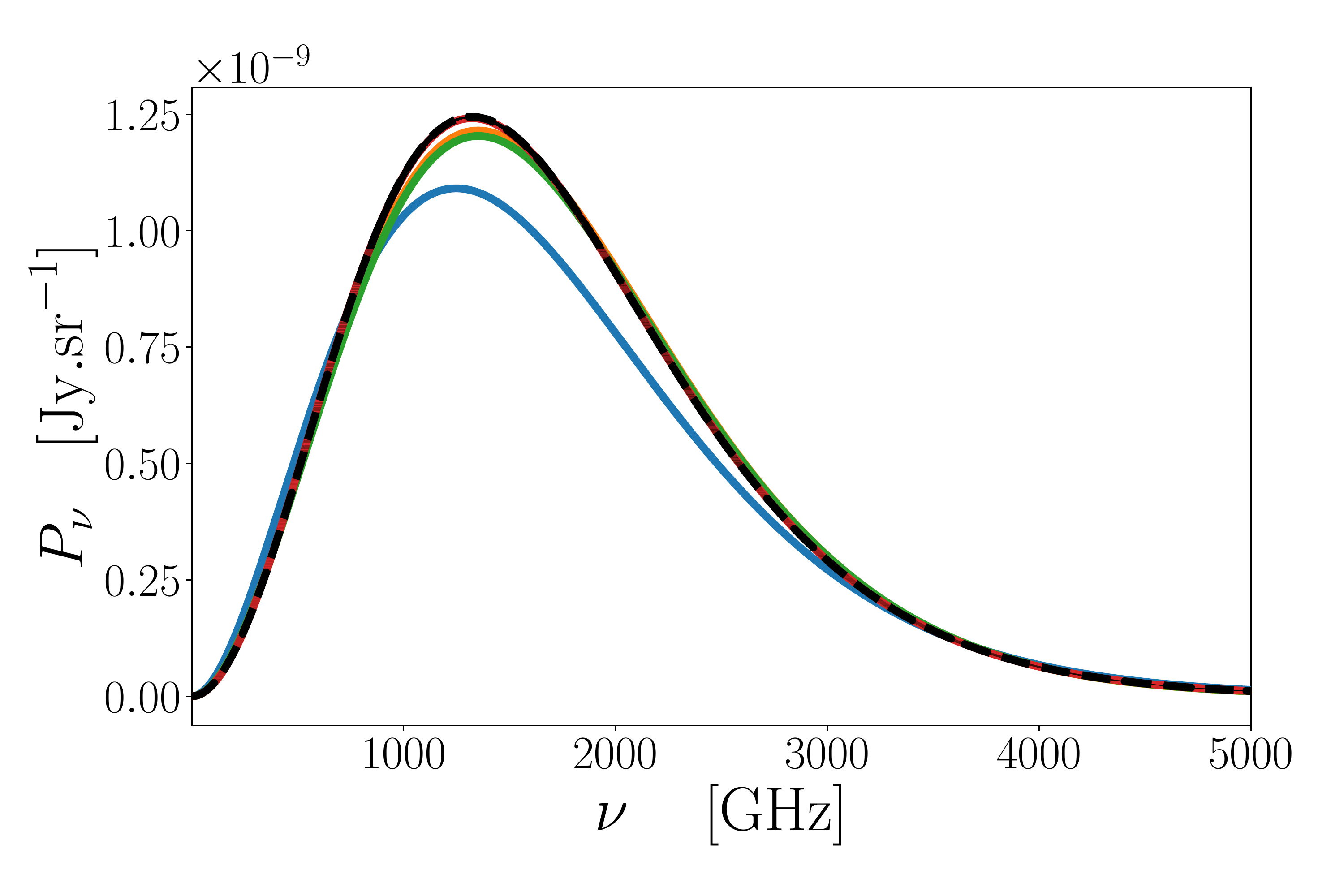}
    \\[1mm]
    \includegraphics[width=0.92\columnwidth]{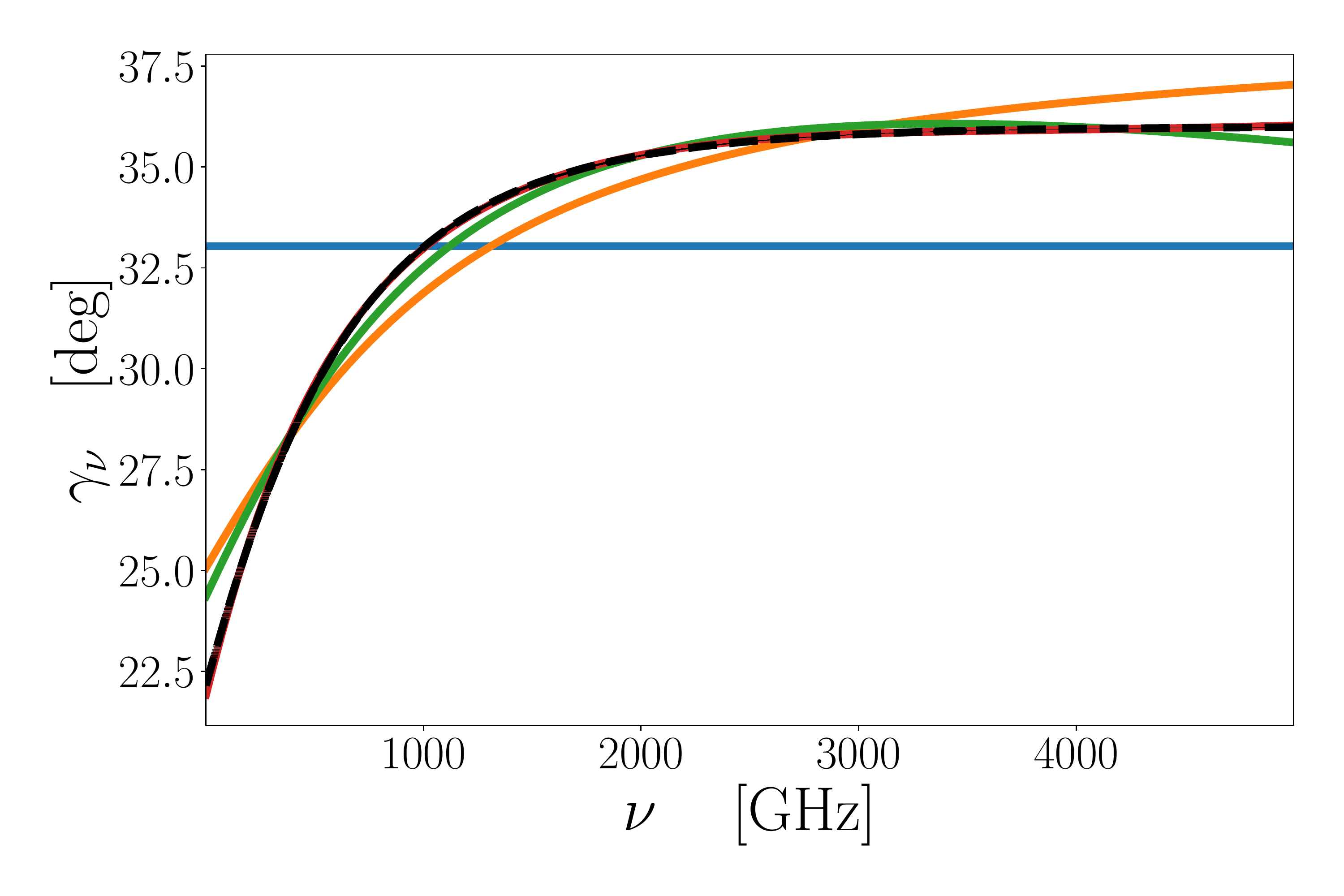}
    \\[1mm]
    \caption{Illustration of  $\spinpol_\nu$ in the complex plane $(Q,U)$  for a sum of two gray-bodies ({\it Upper panel}). Black crosses mark the steps of 1000 GHz on the signal. The values of frequencies are indicated above the crosses in GHz. The corresponding polarized intensity $\polnu$ ({\it Central panel}) and polarization angle $\gamma_\nu$ ({\it Lower panel}).
    %
    % with parameters $A_1=10^6$, $A_2= 10^6$, $\beta_1=1/20$, $\beta_2=1/12$,
    % $2\gamma_1 = 72^{\circ}$,
    % $2\gamma_2 = -90^{\circ}$. 
    % %
    % The best fit using moment expansion is displayed at various orders: leading order (blue), order 1 (orange), order 2 (green) and order 3 (red).
    }
    \label{fig:2GB-P-gamma}
\end{figure}
%--------------------------------------

In Fig.~\ref{fig:2GB-P-gamma}, we fit the above model on a sum of two gray-bodies with parameters $A_1=A_2= 10^6$, $\beta_1=1/20$\,K$^{-1}$, $\beta_2=1/13$\,K$^{-1}$, $2\gamma_1 = 72^{\circ}$, $2\gamma_2 = -90^{\circ}$. To catch the domain on which $\beta_{\rm GB}$ has a maximal impact, we choose $\nu_{\rm max}=5000$\,GHz. 
We note the "loop" trajectory of $\spinpol_\nu$ in the complex plane $(Q,U)$ inherited from the combination of the shape of the black-body SED and the frequency rotation. By definition again, $O(0)$ can only be a straight line and fail to grasp this complexity.
Even if this case again is non-perturbative, we see that the moment expansion up to third order allows us to gradually account for the SED distortions and recover $\polnu$, the polarized mixing inducing a highly nontrivial rotation of the polarization angle. 

%\subsubsection{Perturbative regime}
%--------------------------------------
Like in the power-law example, in the perturbative regime we can obtain a more general expression for the leading order terms. Assuming that $|\mathcal{W}_0|\gg |\mathcal{W}_1^{\betaGB}|$, we can again use the split $\Delta \betaGBb=\mathcal{W}_1^{\betaGB}/\mathcal{W}_0=a_{\Delta \betaGBb}+\i b_{\Delta \betaGBb}$ into real and imaginary parts. 
With this, we can then write
%--------------------------------------
\vspace{-2mm}
\begin{align}
\langle\spinpol^{\rm GB}_\nu\rangle 
%&\approx \bar{A} \hatpol^{\rm GB}_\nu(\bar{T})\times \Bigg\{\mathcal{W}_0 
%+ \mathcal{W}^{\betaGB}_1\,Y^{\rm GB}_1(x)  \Bigg\}
%\nonumber\\
&\approx \bar{A} \mathcal{W}_0\,\hatpol^{\rm GB}_\nu(\bar{T})\times \Bigg\{ 
1 + \Delta \betaGBb\,Y^{\rm GB}_1(x)  \Bigg\}
\nonumber\\
&
\approx \bar{A} \mathcal{W}_0\,\hatpol^{\rm GB}_\nu(\tilde{T})
\end{align}
%--------------------------------------
with $\tilde{T}=1/(\betaGBb+\Delta \betaGBb)$. The leading order SED term, $\hatpol^{\rm GB}_\nu(\tilde{T})$, then depends on the function
%--------------------------------------
\begin{subequations}\label{eq:perturbativepivotGB}
\vspace{-2mm}
\begin{align}
\frac{1}{\expf{\frac{h\nu}{k\tilde{T}}}-1}&=
\frac{1}{\expf{x_{\rm R}+\i x_{\rm I}}-1}
=\frac{
\expf{2\i\Delta\gamma^{\rm GB}_\nu}}
{\sqrt{(\expf{x_{\rm R}}-1)^2+2\expf{x_{\rm R}}\,[1-\cos(x_{\rm I})]}}
\\
\Delta\gamma^{\rm GB}_\nu&=\frac{1}{2}\,\tan^{-1}\left(\frac{\expf{x_{\rm R}}\sin(x_{\rm I})}{\expf{x_{\rm R}}\cos(x_{\rm I})-1}\right)
\end{align}
\end{subequations}
%--------------------------------------
with $x_{\rm R}=h\nu\,(\betaGBb+a_{\Delta\betaGBb})/k$ and $x_{\rm I}=h\nu\,b_{\Delta\betaGBb}/k$. One can see that polarized mixing leads to an imaginary photon chemical potential, $\mu=\i x_{\rm I}$. This causes a frequency-dependent rotation of the polarization plane and also modifications to the SED. At high frequencies, one finds $\Delta\gamma_\nu\approx x_{\rm I}/2$, while at low frequencies, one has the constant $\Delta\gamma_\nu\approx \frac{1}{2}\,\tan^{-1}(b_{\Delta \betaGBb}/a_{\Delta\betaGBb})$. 

\subsection{Modified blackbodies}
\label{sec:mBB_pol}
%--------------------------------------
Another highly relevant SED is given by the modified blackbody spectrum.  It is expected to provide a good model for the thermal dust intensity and polarized signal \citep{Planck2014dust,Planck2015dust}. In principle, one should allow for amplitude, temperature and spectral index variations inside each voxel. This provides multiple ways of creating polarization (see Sec.~\ref{sec:comments_pol} for discussion). However, we assume that the voxel polarization is again only given by amplitude variations in the four emission directions. The fundamental voxel SED then reads $\hatpol^{\rm mBB}_\nu \equiv (\nu/\nu_0)^{\betad}\,B_\nu(T)$, such that the single polarization state can be characterized by $\spinpol^{\rm mBB}_\nu\approx A\,\expf{2\i\gamma} (\nu/\nu_0)^{\betad}\,B_\nu(T)$\footnote{This is the standard way to model polarized dust emission locally, with $A= \pi_{\rm dust} \tau_{\rm dust} \cos^2(\Gamma_{\rm dust})$, $\pi_{\rm dust}$ being the polarization fraction, $\tau_{\rm dust}$ the opacity and $\Gamma_{\rm dust}$ the angle between the Galactic magnetic field and the plane of the sky \citep{Draine_2009}. In practice one would have to consider a correlation between $\Gamma_{\rm dust}$ and $\gamma$.}. 
%--------------------------------------
\begin{figure}[t!]
    \centering
    \includegraphics[width=0.92\columnwidth]{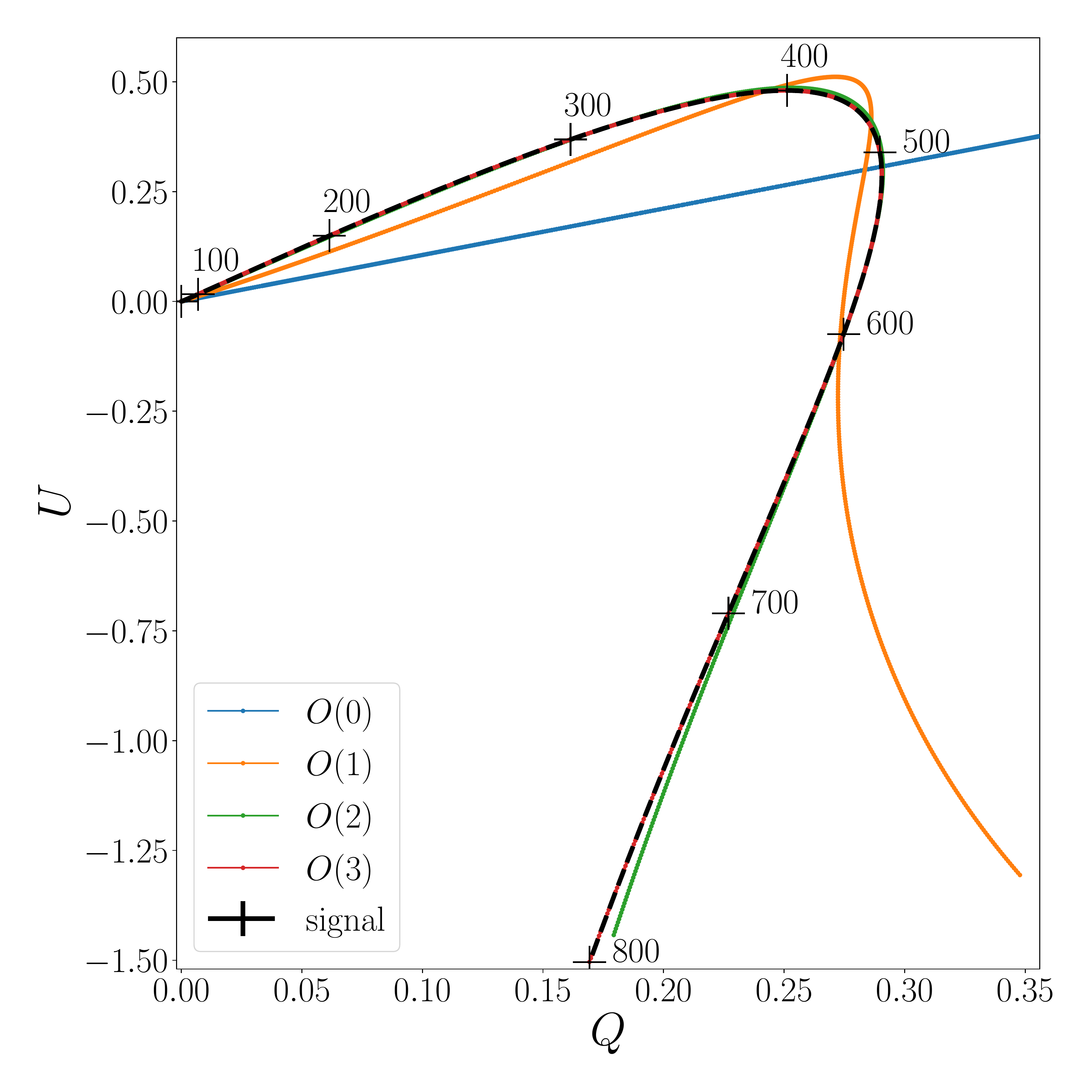}
    \\
    \includegraphics[width=0.92\columnwidth]{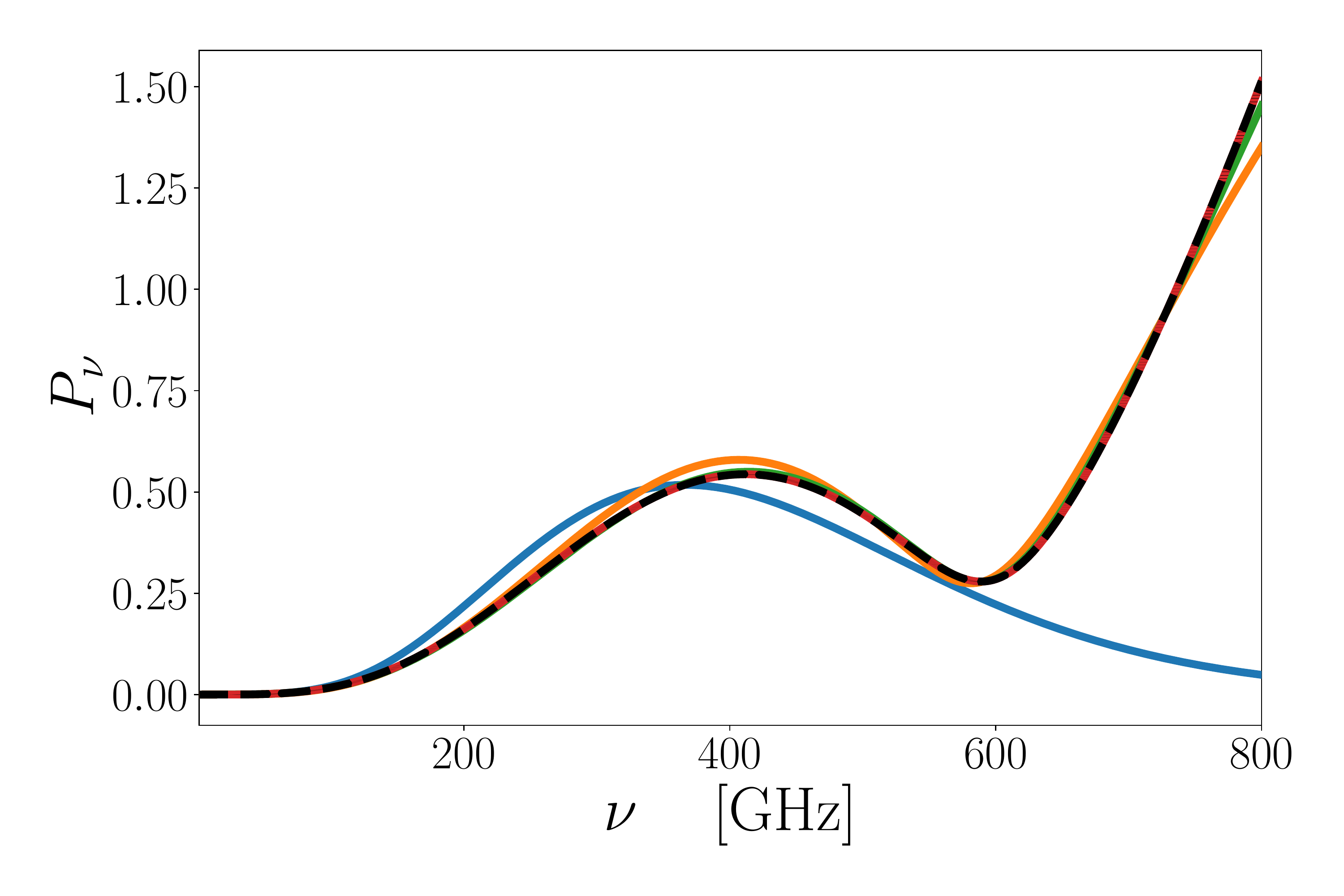}
    \\
    \includegraphics[width=0.92\columnwidth]{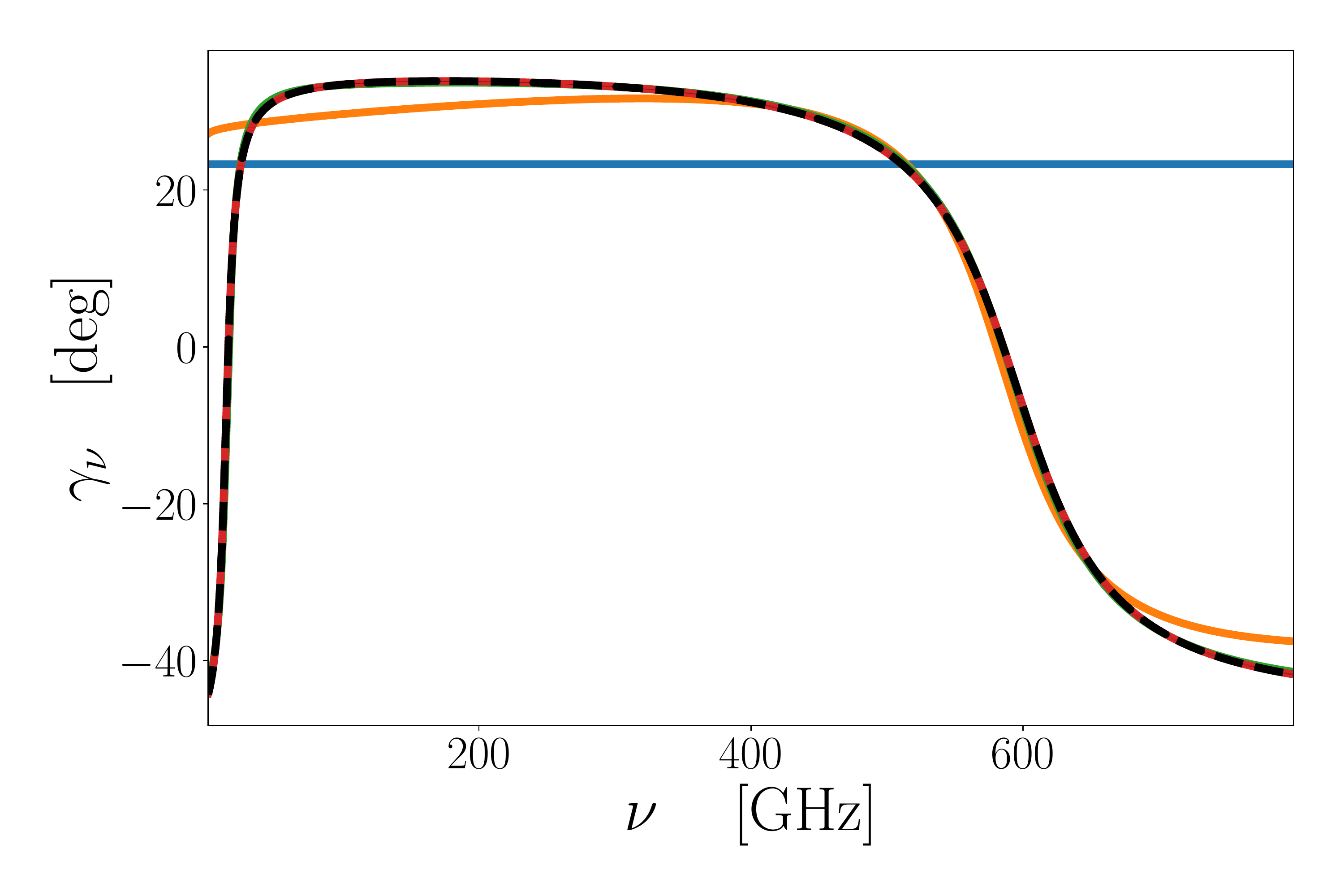}
    \\
    \caption{Illustration of the spinor $\spinpol_\nu$ in the complex plane $(Q,U)$ for a sum of two modified-blackbodies ({\it Upper panel}). Black crosses mark the steps of 100 GHz on the signal. The values of frequencies are indicated above the crosses in GHz. The corresponding polarized intensity $\polnu$ ({\it Central panel}) and polarization angle $\gamma_\nu$ ({\it Lower panel}).
    %
    % with parameters $A_1=1$, $A_2= 1$, $\beta_1=2$, $\beta_2=1$,
    % $T_1=7$\,K, $T_2=70$\,K
    % $2\gamma_1=  72^{\circ}$,
    % $2\gamma_2=  -90^{\circ}$. The best fit using moment expansion is displayed at various orders: leading order (blue), order 1 (orange), order 2 (green) and order 3 (red). {\it Lower panel}: Similar as above for the polarized angle $\gamma_\nu$.
    }
    \label{fig:2MBB-P-gamma}
\end{figure}
%--------------------------------------
Using the results for the power law and gray-body spectra of the previous sections, we then have
%--------------------------------------
\begin{subequations}
\label{eq:mBB_momentExp}
\begin{align}
\langle\spinpol^{\rm mBB}_\nu\rangle 
&=  \polnu^{\rm mBB}(\bar{A},\bar{T}, \betadb)\times \bigg\{\mathcal{W}_0 
+\mathcal{W}^{\betad}_1\ln(\nu/\nu_0)
+\mathcal{W}^{\betaGB}_1 Y^{\rm GB}_1(x)
\nonumber \\ 
+ \frac{1}{2}\mathcal{W}^{\betad^2}_2 & \ln^2(\nu/\nu_0)
+ \mathcal{W}^{\betaGB\betad}_2 \ln(\nu/\nu_0)\,Y^{\rm GB}_1(x)
+\frac{1}{2}\mathcal{W}^{\betaGB^2}_2 Y^{\rm GB}_2(x)
\nonumber \\
&
+\frac{1}{6}\mathcal{W}^{\betad^3}_3 \ln^3(\nu/\nu_0)
+\frac{1}{2}\mathcal{W}^{\betaGB\betad^2}_3 \ln^2(\nu/\nu_0)\,Y^{\rm GB}_1(x)
\\ \nonumber
&
+\frac{1}{2}\mathcal{W}^{\betaGB^2\betad}_3 \ln(\nu/\nu_0)\,Y^{\rm GB}_2(x)
+\frac{1}{6}\mathcal{W}^{\betaGB^3}_3 Y^{\rm GB}_3(x)+ \cdots \bigg\}
\\
\bar{A}&=\langle A\rangle, \quad
\mathcal{W}_0 = \frac{\langle A\,\expf{2\i \gamma}\rangle}{\bar{A}}, \quad
\\
\betaGBb&\equiv\frac{1}{\bar{T}}=\frac{\langle A\,\betaGB\rangle}{\bar{A}},
\quad
\betadb=\frac{\langle A\,\betad\rangle}{\bar{A}}
\\
\mathcal{W}^{\betaGB^\alpha\betad^\delta}_{\alpha+\delta}&=
\frac{\langle A\,\expf{2\i \gamma}\,(\betaGB-\betaGBb)^\alpha\,(\betad-\betadb)^\delta\rangle}{\bar{A}}
\end{align}
\end{subequations}
%--------------------------------------
up to third order. Due to the dimensionality of the problem, the moment representation quickly becomes cumbersome, but can be easily handled using modern computers.

In Fig.~\ref{fig:2MBB-P-gamma}, we applied this expansion on the sum of two normalized\footnote{The two modified blackbodies are here normalized by a reference blackbody at $\nu=\nu_0$, as further discussed in Appendix~\ref{sec:normalization}. As such, $Q$, $U$ and $\polnu$ are unitless.} modified blackbodies of parameters $A_1= A_2= 1$, $\beta_1=2$, $\beta_2=1$,
$T_1=5$\,K, $T_2=70$\,K,
$2\gamma_1=  72^{\circ}$,
$2\gamma_2=  -90^{\circ}$. To simulate the thermal dust signal over the CMB missions frequency ranges and in order to witness the transition between the effect of the power-law factor at low frequencies and the gray-body factor at high frequencies, we choose $\nu_{\rm max} = 800$\,GHz. Accordingly to the \planck{} high frequency bands, we choose $\nu_0=353$\,GHz \citep{planck_2015_overview}. One can see that the different power laws induce strong distortions at low frequencies $\leq 100~{\rm GHz}$ while the temperature induces an additional bending at high frequencies. In this very extreme case, all the moments up to order 3 are required to correctly model the signal. However, even in this nontrivial situation the moment expansion performs extremely well. The leading order fit $O(0)$ interprets a local maximum of the polarized intensity and as the peak of the gray-body spectra, leading to a wrongly small value for the recovered temperature.
In astrophysical situations one expects the central limit theorem to render the examples more moderate, with fewer moments required at a given precision.

%\subsection{Perturbative regime}
%--------------------------------------
In the perturbative regime, using the results from the previous sections, the leading order moment description then is
%--------------------------------------
\begin{subequations}\label{eq:perturbativepivotMBB}
\begin{align}
\langle\spinpol^{\rm mBB}_\nu\rangle 
&\approx
\frac{2 h \nu^3}{c^2}\frac{\bar{A}\, \Omega_0\,(\nu/\nu_0)^{\betadb+a_{\Delta\betadb}}\,\expf{2\i \gamma_\nu}}
{\sqrt{(\expf{x_{\rm R}}-1)^2+2\expf{x_{\rm R}}\,[1-\cos(x_{\rm I})]}}
\\
\gamma_\nu&=\gamma_0+\frac{1}{2}\,b_{\Delta \betadb}\ln(\nu/\nu_0)+\Delta\gamma^{\rm GB}_\nu,
\end{align}
\end{subequations}
%--------------------------------------
with definitions as in the previous section. This expression demonstrates that the rotation of the polarization plane now has two contributions, one from the power-law modulation, one from the temperature terms. The rotation caused by temperature terms is particularly important at high frequencies and can become rapid due to a near linear scaling with $\nu$.

\section{Generalizations of the formalism}
\label{sec:applications2}
%------------------------------------
In this section, we discuss additional averaging processes, covering the spherical harmonic decomposition and beam averaging effects. These cases all naturally lead to a redefinition of the meaning and values of the moments and spectral pivot, but they {\it do not} actually change the structure of the moment representation. For each case, we briefly recap how the problem is treated in intensity before generalizing to spin moments.

\subsection{Generalization to spherical harmonics}
%------------------------------------
As it was done for intensity in \cite{Chluba}, one can immediately generalize the spin-moment expansion in harmonic space. For that we have to leave the above restriction of considering a single line of sight $\vecn$ and consider the intensities as fields over the celestial sphere. 
Using the moment expansion formalism at the power spectra level is especially useful for component separation on large sky fractions as it has been already shown for \planck{} data \citep{Mangilli}, \so{} telescope \citep{Azzoni} and \lb{} \citep{RemazeillesmomentsILC,Vacher21}.

\subsubsection{In intensity}
%------------------------------------
While in the above, we described the signal along a given line of sight $\vecn$, we now turn ourselves to averages between different lines-of-sights across sky patches. The average intensity, $\langle I_\nu\rangle$, which generally depends on the line-of-sight moments and pivots of the SED expansion in the direction $\vecn$, is a scalar field on the $S^2$-manifold. As such, it can be expanded on the orthogonal basis of the spherical harmonic functions $Y_{\ell\,m}(\vecn)$.
%as:
% %--------------------
% \begin{equation}
% I_\nu(\vecn) = \sum_{\ell=1}^\infty \sum_{m=-\ell}^\ell (I_\nu)_{\ell\,m} Y_{\ell\,m}(\vecn)
% \end{equation}
% %--------------------
% Where $(I_\nu){\ell\,m} \in \mathbb{C}$ and $Y_{\ell\,m}: S^2 \to \mathbb{C}$.
%
We shall denote the spherical harmonic coefficients of a quantity $X$ as 
%--------------------
\begin{align}
(X)_{\ell m}=\int {Y}^{*}_{\ell m}(\vecn) \,X(\vecn)\,{\rm d}^2\vecn.
\end{align}
%--------------------
The spherical harmonic coefficients of the average intensity SED can then be expressed as:
%--------------------
\begin{align}
\label{eq:I_av_sky}
\langle I_\nu \rangle_{\ell\,m}&\equiv
\langle(I_\nu)_{\ell\,m}\rangle
=
\int {Y}^{*}_{\ell m}(\vecn) \,\prob(\vec{p}, \vecn) \,\hat{I}_\nu(\vec{p})  \,{\rm d}^N p\,{\rm d}^2\vecn
\nonumber\\
&=
\int \prob_{\ell m}(\vec{p}) \,\hat{I}_\nu(\vec{p})  \,{\rm d}^N p, 
\end{align}
%--------------------
where $\prob_{\ell m}(\vec{p})$ is the harmonic coefficient of the parameter distribution, $\prob(\vec{p}, \vecn)$.
Just as the single line-of-sight average, this superposition will introduce some additional mixing between intensities with different spectral parameters $\vec{p}(\vecn)$ and introduces extra distortions. The spherical harmonic functions introduce a reweighting of the parameter distribution, which essentially implies that all moments can be directly computed using the harmonic coefficients of the parameter distribution, $(P)_{\ell m}(\vec{p})$.

Expanding $\hat{I}_\nu(\vec{p})$ in Eq.~\eqref{eq:I_av_sky} and following the same steps introduced in Sec.~\ref{sec:momintensity}, one can derive the moment expansion in harmonic space having the exact same structure as Eq.~\eqref{eq:moments-intensity}:
%--------------------
\begin{align}
\label{eq:moments-ell}
    \langle I_\nu\rangle_{\ell\,m} &=  \delta_{\ell0}\,I_\nu(\barA,\barpv) + \sum_j(\omega_1)^{p_j}_{\ell\,m} \partial_{\barp_j} I_\nu(\barA,\barpv)
    \nonumber \\
    &\quad+ \frac{1}{2}\sum_{j,k}(\omega_2)_{\ell\,m}^{p_j p_k}\partial_{\barp_j}\partial_{\barp_k}I_\nu(\barA,\barpv)
     \nonumber\\
    &\qquad+ \frac{1}{3!}\sum_{j,k,l}(\omega_3)_{\ell\,m}^{p_j p_k p_l}\partial_{\barp_j}\partial_{\barp_k}\partial_{\barp_l}I_\nu(\barA,\barpv) + \dots \,,
\end{align}
%--------------------
where $\delta_{ij}$ is the Kronecker-$\delta$ and we introduced the (complex-valued) moment multipoles
%--------------------
\begin{align}
    (\omega_\alpha)^{p_j\dots p_l}_{\ell m}&= \frac{\langle A(p_j - \bar{p}_j)\dots(p_l - \bar{p}_l)\rangle_{\ell\,m}}{\barA}.
\end{align}
%--------------------
In Eq.~\eqref{eq:moments-ell}, we  used the average of all the parameters across the sky to define the average SED amplitude and pivot\footnote{Note that in \citet{Mangilli} and its follow-up papers (e.g. \cite{Vacher21}), the existence of a scale dependent pivot $\barppolv(\ell)$ in the harmonic space level has been assumed. Proving or discussing formally this assumption is left for future work.}
%--------------------
\begin{align}
\barA&=\int \prob(\vec{p}, \vecn) \,{\rm d}^N p\,
\frac{{\rm d}^2\vecn}{4\pi} \quad{\rm and}\quad
\barpv=\frac{\int \prob(\vec{p}, \vecn) \,\vec{p}\,{\rm d}^N p\,
\frac{{\rm d}^2\vecn}{4\pi}}{\int \prob(\vec{p}, \vecn) \,{\rm d}^N p\,
\frac{{\rm d}^2\vecn}{4\pi}}.
\end{align}
%--------------------
This choice might be best suited for convergence (canceling the first order moment of the intensity expansion over the whole sky) when done in real space.
%--------------------
This expansion is straightforward to generalize to angular power-spectra and cross-frequency power spectra \citep{Mangilli} and is what has so far been used to describe $B$-modes signal.

\subsubsection{In Polarization}

The averaged polarized signal $\langle \spinpol_\nu(\vec{p}(\vecn))\rangle=\langle\spinpol_\nu(\vecn)\rangle$ is now a section of the spin-2 bundle on the $S^2$-manifold. It can be expanded on the orthogonal basis of the spin-2 weighted spherical harmonics ${_{\pm 2}}Y_{\ell m}$ \citep{NewmannPenrose1966,Goldberg1967}. One can evaluate the spherical-harmonic coefficients of a general frequency dependent spinor field $_{\pm 2}\mathcal{X}_\nu$ for $\ell \geq 2$  as\footnote{The spin-weighted harmonics function ${_{s}}Y_{\ell\,m}$ are defined for $\ell\geq |s|$.} 
%---------------------------
\begin{equation}
_{\pm 2}(\mathcal{X}_\nu)_{\ell\,m} = \int {_{\pm 2} Y}^{*}_{\ell m}(\vecn) \,\mathcal{X}_\nu(\vecn)\, {\rm d}^2\vecn
\label{eq:Inuellm2}
\end{equation}
%---------------------------
For convenience, we then also define the harmonic expansion of the line-of-sight average of the polarization field as %---------------------------
\begin{align}
\label{eq:I_av_sky_pol_gen}
_{\pm 2}\langle \spinpol_\nu\rangle_{\ell\,m}
&= \int {_{\pm 2} Y}^{*}_{\ell m}(\vecn)\,
\prob(\vec{p},\gamma, \vecn) \,\spinpol_\nu(\vec{p},\gamma)  \,{\rm d}^N p \,{\rm d}\gamma \, {\rm d}^2\vecn
\nonumber\\
&=
\int {_{\pm 2}}\prob_{\ell m}(\vec{p},\gamma) \,\spinpol_\nu(\vec{p},\gamma)  \,{\rm d}^N p \,{\rm d}\gamma,
\end{align}
%---------------------------
where we define the harmonic coefficient of the parameter distribution function, $\prob(\vec{p},\gamma, \vecn)$, as ${_{\pm 2}}\prob_{\ell m}(\vec{p},\gamma)$.
When applied to the spin-moment expansion of the polarization field, we then obtain 
%---------------------------
\begin{align}
\label{eq:moments-ell2}
    {_{\pm 2}}\langle \spinpol_\nu\rangle_{\ell\,m} &=  
    \sum_j{_{\pm 2}}(\mathcal{W}_1^{p_j})_{\ell\,m}\, \partial_{\barp_j} \polnu(\barA,\barpv)
    \nonumber \\
    &\quad+ \frac{1}{2}\sum_{j,k} {_{\pm 2}}(\mathcal{W}_2^{p_j p_k})_{\ell\,m}\,\partial_{\barp_j}\partial_{\barp_k}\polnu(\barA,\barpv)
     \\ \nonumber
    &\qquad+ \frac{1}{3!}\sum_{j,k,l}{_{\pm 2}}(\mathcal{W}_3^{p_j p_k p_l})_{\ell\,m}\,\partial_{\barp_j}\partial_{\barp_k}\partial_{\barp_l}\polnu(\barA,\barpv) + \dots \,,
\end{align}
%---------------------------
with the spin-moment multipoles
%---------------------------
\begin{align}
    {_{\pm 2}}(\mathcal{W}_\alpha^{p_j\dots p_l})_{\ell,m}&= \frac{{_{\pm 2}}\langle A\expf{2\i \gamma}(p_j - \bar{p}_j)\dots(p_l - \bar{p}_l)\rangle_{\ell\,m}}{\barA}.
\end{align}
%---------------------------
Since this expansion is not defined for $\ell\leq2$, one does not have any leading order (monopole) term. Nevertheless, the spectral pivot is again defined by the average over the full sky,  
%--------------------
\begin{subequations}
\begin{align}
\barA&=\int \prob(\vec{p}, \gamma, \vecn) \,{\rm d}^N p\,{\rm d}\gamma\,
\frac{{\rm d}^2\vecn}{4\pi}
\\
\barpv&=\frac{\int \prob(\vec{p}, \gamma, \vecn) \,\vec{p}\,{\rm d}^N p\,{\rm d}\gamma\,
\frac{{\rm d}^2\vecn}{4\pi}}{\int \prob(\vec{p}, \gamma, \vecn) \,{\rm d}^N p\,
\frac{{\rm d}^2\vecn}{4\pi}}
\end{align}
\end{subequations}
%--------------------
which is the only physically motivated choice. It now has to include the average over $\gamma$. 

\subsubsection{{\it E}- and {\it B}-modes}
%--------------------
From the harmonic coefficients of the polarized spinor, one can then obtain the $E$- and $B$-mode coefficients of the polarized SED as \citep{Zaldarriaga1997,Kamionkowski}:
%--------------------
\begin{subequations}
\label{eq:defE-B}
\begin{align}
& \langle\spinpol_\nu\rangle_{\ell m}^E=-\frac{1}{2}\left[ {_{+2}}\langle\spinpol_\nu\rangle_{\ell m} + {_{-2}}\langle\spinpol_\nu^*\rangle_{\ell m} \right]\\
& \langle\spinpol_\nu\rangle_{\ell m}^B=-\frac{1}{2i} \left[ {_{+2}}\langle\spinpol_\nu\rangle_{\ell m} - {_{-2}}\langle\spinpol_\nu^*\rangle_{\ell m} \right].
\end{align}
\end{subequations}
%--------------------
One can then expand both $\spinpol_\nu$ and its complex conjugate using Eq.~\eqref{eq:moments-ell2} and insert them into Eq.~\eqref{eq:defE-B}. Going from the expansion of the previous section to the $E$- and $B$-modes adds \emph{no extra averaging} effect. For each mode, we get a moment expansion with the same structure as Eq.~\eqref{eq:moments-ell2} with spin moments:
%--------------------
\begin{subequations}
\label{eq:momE-B}
\begin{align}
& (\mathcal{W}^{p_j\dots p_l}_\alpha)^{E}_{\ell m}=-\frac{1}{2}\left[ {_{+ 2}}(\mathcal{W}^{p_j\dots p_l}_\alpha)_{\ell m} + {_{- 2}}({\mathcal{W}^{p_j\dots p_l}_\alpha}^*)_{\ell m}\right],\\
& (\mathcal{W}^{p_j\dots p_l}_\alpha)^{B}_{\ell m}=-\frac{1}{2i} \left[ {_{+ 2}}(\mathcal{W}^{p_j\dots p_l}_\alpha)_{\ell m} - {_{- 2}}({\mathcal{W}^{p_j\dots p_l}_\alpha}^*)_{\ell m} \right].
\end{align}
\end{subequations}
%--------------------
Since this new expansion is simply derived from the one of $\spinpol_\nu$, it inherits its amplitude and pivot defined with the $\gamma$-weighted full sky averages.
This justifies the approximation used in previous studies where the $B$-mode signal was treated as an intensity at the map  \citep[e.g.,][]{RemazeillesmomentsILC} or power-spectra level \citep[e.g.,][]{Azzoni,Vacher21} %with the additional subtlety that one should not use any leading order.
We leave a detailed discussion on the subtleties of $E$- and $B$- modes and power-spectra generalizations for future work.

\subsection{Averages inside the beam and bandpass effects}
\label{sec:beams}
%-------------------------------------------------
Introducing the general instrumental transfer function $W(\nu,\gamma, \vecn)$ will add additional mixing between the lines of sight but also new nontrivial spectral dependencies from the mixing in frequency. In practice, this very general $W$ function could also account for some effects due to intensity to polarization leakage. 

The total polarized signal in a frequency band of width $\Delta\nu$ at average frequency $\nu_c$ and within the spatial support of the beam, $\Omega$, centered in an average line of sight $\vecn_c$ then becomes:
%--------------------
\begin{align}
\label{eq:I_av_sky_beam}
\langle \spinpol_{\nu}\rangle_{W}
=\int_\Omega & \prob(\vec{p},\gamma, \vecn) \,W(\nu,\gamma, \vecn) \,\hatpol_\nu(\vec{p})\,\expf{2\i \gamma} \,{\rm d}^N p  \,{\rm d}\gamma \, d^2\vecn  \,{\rm d}\nu. 
\end{align}
%--------------------
Here, $\langle \spinpol_{\nu}\rangle_{W}$ is now a function of $\nu_c$ and the corresponding moments and pivots in the average direction $\vecn_c$.

Due to the $\nu$ integration in the above expression, it is not possible to simply factor the spectral shapes in the Taylor expansion of $\hatpol_\nu$ out of the integral as we did before \citep{Chluba}. The band-pass function of the instrument therefore couples different spatial regions with different spectral forms. One can still generalize the spin-moment expansion to account for this extra averaging. The new expansion becomes:
%--------------------
\begin{align}
\label{eq:spin-mom-beam}
    &\langle \spinpol_{\nu}\rangle_{W}
    =  \mathcal{W}_0\,\langle\hatpol_\nu(\barA,\barppolv)\rangle_{W}
    + \sum_j \mathcal{W}_1^{p_j} \langle \partial_{\barppol_j}\hatpol_\nu(\barA,\barppolv)\rangle_{W}
    \nonumber \\  
    &\qquad\quad
    + \frac{1}{2}\sum_{j,k} \mathcal{W}_2^{p_j p_k} \langle
    \partial_{\barppol_j}\partial_{\barppol_k}\hatpol_\nu(\barA,\barppolv)\rangle_{W}
    \\ \nonumber
    &\qquad\qquad
    +\frac{1}{\alpha!} \sum_{j,\dots,l} \mathcal{W}_\alpha^{p_j\dots p_l} 
    \langle\partial_{\barppol_j}\dots\partial_{\barppol_l}\hatpol_\nu(\barA,\barppolv)\rangle_{W}+ \dots,
\end{align}
%--------------------
with the new averages defined as 
%--------------------
\begin{equation}
\langle \mathcal{X} \rangle_{W} = \int  \prob(\vec{p}, \gamma,\vecn)\, W(\nu,\gamma, \vecn) \, \mathcal{X} \,{\rm d}^N p \,{\rm d}\gamma \,d^2\vecn \,{\rm d}\nu. 
\end{equation}
%--------------------
The pivot is chosen to be the same as in intensity:
%--------------------------
\begin{equation}
\label{eq:pivot_beam}
\barppol_j  = \frac{\langle A p_j \partial_{\barppol_j}\polnu(\barpv)\rangle_{W}}{\langle A\partial_{\barppol_j}\polnu(\barpv)\rangle_{W}}
\end{equation}
%--------------------------
and the spin moments become:
%--------------------
\begin{equation}
\label{eq:spin-moments_beam}
   \mathcal{W}_\alpha^{p_j\dots p_l}
    =\frac{\left<A\,\expf{2\i\gamma} (p_j-\barppol_j)\dots(p_l-\barppol_l)\partial_{\barppol_j \dots \barppol_l }\hatpol_\nu(\vec{p})\right>_{W}}{\left< A \partial_{\barppol_j \dots \barppol_l }\hatpol_\nu(\vec{p})\right>_{W}}.
\end{equation}
%--------------------------
These expressions assume that the various SED derivative averages do not vanish, however, the conclusions are not affected.

Generally, one cannot disentangle the mixing due to the physics of the source and the one due to the instrumental response, and in practice this moment expansion can become awfully complicated. A frequent assumption is to introduce a factorization of $W$ as a band-pass term $F$ times  a spatial polarized beam shape $\mathfrak{B}$ as $W(\nu, \vecn)=\mathfrak{B}(\gamma, \vecn)F(\nu)$ see e.g. \cite{PlanckIX-2013}. In this case, it is possible to split the integral of Eq.~\eqref{eq:I_av_sky_beam} to factorize the frequency dependent terms, allowing the vanishing of the spectral averages such that the expressions for the pivot and the spin moments are similar to the ones derived for the single line-of-sight case, extending only the averages to the multiple lines of sight included in $\mathfrak{B}$. The result remains nontrivial since the spectral dependence of the moments need to be computed through the potentially complicated integral of $F(\nu)\partial_{p_j\dots p_k}\hatpol_\nu$ over $\Delta\nu$. The expansion can also be different in each band of the instrument, since they can have a different instrumental response $W$. Treating these cases in more detail is beyond the scope of this work.

\section{Additional aspects}
\label{sec:discussion}

\subsection{Effect of voxel-level SED variations}
\label{sec:SEDvoxelvar}
%--------------------------
As stressed in Sect.~\ref{sec:comments_pol}, in most of this work we considered that inside each voxel, the net linear polarization have a single SED of which $Q$ and $U$ are the projections. The average signal then inherits a frequency-dependent polarization angle solely from polarized mixing across voxels. Among our examples in Sect.~\ref{sec:applications}, the only exception was the mixing of blackbody spectra, where the polarization degree was caused by variations of the main spectral parameter, the blackbody temperature.

{\it Can we extend the moment formalism to include voxel-level SED mixing caused by variations of the spectral parameters?}
Let us use the power-law SED as the example. In addition to the weight parameter variations in the different directions, we now also have variations of the spectral index, $\beta$. Considering only $Q_\nu$, from the discussion in Sect.~\ref{sec:comments_pol} we then have %--------------------------------------
\begin{align}
Q_\nu=&\frac{A_\parallel\,\hatpol^{\rm PL}_\nu(\beta_\parallel)-A_\perp\,\hatpol^{\rm PL}_\nu(\beta_\perp)}{2}
=\frac{(A_\parallel-A_\perp)}{2}
\,\frac{\hatpol^{\rm PL}_\nu(\beta_\parallel)
+\hatpol^{\rm PL}_\nu(\beta_\perp)}{2}
\nonumber\\
&\qquad\qquad\qquad+\frac{A_\parallel+A_\perp}{2}
\,\frac{\hatpol^{\rm PL}_\nu(\beta_\parallel)-\hatpol^{\rm PL}_\nu(\beta_\perp)}{2}.
\end{align}
%--------------------------------------
A similar expression follows for $U_\nu$, with the relevant parameters $A_\times, \beta_\times, A_\otimes, \beta_\otimes$. Let us again expand the fundamental SEDs around some average index $\bar{\beta}$ inside each voxel.\footnote{This in fact is $\bar{\beta}_{\rm v}=(A_\parallel\beta_\parallel
+A_\perp \beta_\perp+A_\times \beta_\times+A_\otimes \beta_\otimes)/(A_\parallel+A_\perp+A_\times+A_\otimes)$.} This then yields
%--------------------------------------
\begin{align}
Q_\nu
=&\frac{(A_\parallel-A_\perp)}{2}
\,\hatpol^{\rm PL}_\nu(\bar{\beta})
\left[1+\sum_{k=1}^\infty
\frac{\Delta\beta_\parallel^k+\Delta\beta^k_\perp}{2k!}
\ln^k(\nu/\nu_0)\right]
\\ \nonumber
&\qquad \qquad
+\frac{A_\parallel+A_\perp}{2}
\,\hatpol^{\rm PL}_\nu(\bar{\beta})
\left[\sum_{k=1}^\infty
\frac{\Delta\beta_\parallel^k-\Delta\beta^k_\perp}{2k!}
\ln^k(\nu/\nu_0)\right]
\\\nonumber
&= \hatpol^{\rm PL}_\nu(\bar{\beta})\left(
\frac{(A_\parallel-A_\perp)}{2}
\,
+
\,
\sum_{k=1}^\infty
\frac{A_\parallel\Delta\beta_\parallel^k-A_\perp\Delta\beta^k_\perp}{2k!}\ln^k\fracnu\right)
\end{align}
%--------------------------------------
with $\Delta\beta_k=\beta_k-\bar{\beta}$. We already considered the first term for which spectral index variations between voxels lead to the moment expansion. In addition to these voxel-to-voxel variations one now obtains additional terms that are related to variations within the voxel. All these will cause a redefinition of the SED parameter distribution, however, no {\it new} spectral shapes are introduced and hence the moment method equally describes both averaging processes, after the moments are reinterpreted. However, to explicitly make the link to the underlying spectral parameter distributions along the line of sight and within each voxel involves a more complicated description, which we do not provide here.

We close by remarking that, if even the fundamental SEDs in each of the directions differ, then one should simply perform two independent moment expansions accounting for the independent types of SEDS individually. This will quickly increase the number of moments that are required to describe the complexity of the polarized field, but it should nevertheless work even if the physical properties are not independent. The main hope then is that line-of-sight averaging effects reduce the dimensionality of the problem to a manageable level. A more detailed discussion is, however, beyond the scope of this work.

\subsection{Frequency dependence of the polarization angle without mixing: the case of Faraday rotation}
%--------------------------
In the presence of magnetic fields, photons experience Faraday rotation while they propagate through the interstellar medium \citep[e.g.,][]{Heald-FR}. The polarization angle for one voxel at affine parameter $s$ then changes as
%--------------------------
\begin{align}
\gamma_\nu &\approx \gamma + \Gamma f_\nu,
\end{align}
%--------------------------
where $f_\nu = \nu^{-2}$ and $\Gamma(s) \propto \int_0^s n_e(s') B(s') \,{\rm d}s'$ is an integral along the line of sight, with the electron density $n_e(s)$ and the component of the magnetic field in the direction of propagation, $B(s)$. Averaging over various emission points experiencing Faraday rotation adds some extra complications to the moment expansion in a way that is truly unique to polarization. For simplification let us consider the average over a single line of sight. The averaged spinor now becomes:
%--------------------------
\begin{align}
\label{eq:P_av_los_FR}
\langle \spinpol_\nu\rangle_{\rm FR}
&=\int \prob(\vec{p}, \gamma,\Gamma, \vecn) \,\hatpol_\nu(\vec{p})\,\expf{2\i\gamma}\,\expf{2\i\Gamma f_\nu} \,{\rm d}^N p\,{\rm d}\gamma\, {\rm d}\Gamma,
\end{align}
%--------------------------
where we added the level of Faraday rotation as another parameter to the distribution.
Taylor expanding the polarized intensity, one can define the spin-moment expansion using the moments
%--------------------------
\begin{align}
    \mathcal{W}_{\alpha,\rm FR}^{p_j\dots p_l}&= \frac{\langle A\,\expf{2\i \gamma}\,\expf{2\i \Gamma f_\nu}\,(p_j - \bar{p}_j)\dots(p_l - \bar{p}_l)\rangle_{\rm FR}}{\barA_{\rm FR}}.
\end{align}
%--------------------------
with $\barA_{\rm FR}=\langle A \rangle_{\rm FR}$.
The spin moments now become highly nontrivial and generally frequency dependent. 

As discussed in Sec.~\ref{sec:depolarization}, under simplifying assumptions
one can assume that the emission processes themselves (quantified by the spectral parameters $\vec{p}$) are de-correlated from the Faraday rotation experienced by light on its way along the line of sight. This would allow us to write $\prob(\vec{p}, \gamma,\Gamma, \vecn) \approx \prob(\vec{p}, \gamma, \vecn)\,\prob(\Gamma, \vecn)$. %
Physically, this is indeed well-motivated unless the emission process in one voxel knows about the structure of the magnetic field in another (more distant) voxel. The Faraday rotation and the SED averaging in the moments then becomes separable:
%--------------------------
\begin{align}
    \mathcal{W}_{\alpha,\rm FR}^{p_j\dots p_l}
    &\approx 
    %\frac{\langle A\expf{2\i \gamma}(p_j - \bar{p}_j)\dots(p_l - \bar{p}_l)\rangle\,\langle \expf{2\i \Gamma f_\nu}\rangle}{\barA_{\rm FR}}
    %\nonumber\\
    %&=
    \frac{\int \prob(\vec{p}, \gamma, \vecn) \,\hatpol_\nu(\vec{p})\,\expf{2\i\gamma}\,{\rm d}^N p\,{\rm d}\gamma}{\int \prob(\vec{p}, \gamma, \vecn) \,{\rm d}^N p\,{\rm d}\gamma}
    \,\frac{\int \prob(\Gamma, \vecn) \,\expf{2\i \Gamma f_\nu}\,{\rm d}\Gamma}{\int \prob(\Gamma, \vecn) \,{\rm d}\Gamma}
    \nonumber\\[1mm]
    &= \mathcal{W}_\alpha^{p_j\dots p_l}\,\frac{\int \prob(\Gamma, \vecn) \,\expf{2\i \Gamma f_\nu}\,{\rm d}\Gamma}{\int \prob(\Gamma, \vecn) \,{\rm d}\Gamma}.
\end{align}
%--------------------------
This expression shows that no new spectral mixing occurs in this case, as all moments are multiplied by the same frequency-dependent factor. The average Faraday rotation coefficient is simply given by
%--------------------------------------
\begin{align}
\bar{\Gamma}(\vecn)=\frac{\left<A \Gamma\right>_{\rm FR}}{\left<A\right>_{\rm FR}}
\approx\frac{\int \prob(\Gamma, \vecn) \,\Gamma\,{\rm d}\Gamma}{\int \prob(\Gamma, \vecn) \,{\rm d}\Gamma}.
\end{align}
%--------------------------------------
Using this as a pivot, we can then write series
%--------------------------------------
\begin{align}
\label{eq:FR_pol_angle_perturb}
\frac{\left<A\expf{2\i\Gamma f_\nu}\right>_{\rm FR}}{\barA_{\rm FR}}
&=\expf{2\i\bar{\Gamma}\,f_\nu}\left[1+\sum_{k=1}^\infty \frac{(2\i\,f_\nu)^k}{k!} \,\frac{\left<A(\Gamma-\bar{\Gamma})^k\right>_{\rm FR}}{\barA_{\rm FR}}\right],
\end{align}
%--------------------------------------
illustrating how a complicated frequency structure can be created from higher order moments of $\Gamma-\bar{\Gamma}$. Given that these depend on powers of $f_\nu$, one can in principle determine these moments observationally.
Finally, if the simple factorization of the parameter distribution function is not possible, one can write
%--------------------------
\begin{align}
&\mathcal{W}_{\alpha,\rm FR}^{p_j\dots p_l}
=
\expf{2\i\bar{\Gamma}\,f_\nu}
\Bigg[
\mathcal{W}_\alpha^{p_j\dots p_l}
\\ \nonumber
&
\left.
\qquad+\sum_{k=1}^\infty \frac{(2\i\,f_\nu)^k}{k!} \,\frac{\left<A\,\expf{2\i \gamma}\,(p_j - \bar{p}_j)\dots(p_l - \bar{p}_l)\,(\Gamma-\bar{\Gamma})^k\right>_{\rm FR}}{\barA_{\rm FR}}\right].
\end{align}
%--------------------------
This generally introduces a complicated voxel SED reweighting by factors of $(f_\nu)^k$ and hence new SED shapes that in principle can all be separately accounted for. As is understood, the number of variables quickly becomes unmanagable unless highly perturbative situations are encountered. A more detailed discussion is, however, beyond the scope of this work.

% \subsection{Comparison with literature \jenscomment{remove...}}
% %--------------------------
% In their final form, the spin-moments are equivalent to the Delta-map method introduced for component separation in \cite{moment_polar} but at first order only and without proper justification.

% A word on possible extensions of MILC? \cite{MILC,Remazeilles_etal_2016,RemazeillesmomentsILC}

% We justified for the first time the assumption that $B$-modes spectra could be treated as intensity, as done in \cite{RemazeillesmomentsILC,Mangilli,Azzoni,Vacher21}.

\section{Conclusion}
\label{sec:conclusion}
%--------------------------
In the present work, we introduced the spin moments, the natural generalization of the intensity moment expansion introduced in \citet{Chluba} to polarized signals. 
We developed the formalism from basic principles, showing that the moments are promoted to spin-2 complex coefficients that can be expressed in term of the SEDs parameter distribution [see e.g. Eq.~\eqref{eq:spin-moments}]. 
%
% In all the averaging processes considered, these coefficients can be expressed in term of the SEDs parameter distribution (see Eq.~\eqref{eq:moments}).
%
% We displayed the full expansion at all orders and explored scenarios of increasing complexity, considering several canonical SED examples and extending the formalism to all kind of averaging processes.

Thinking about the spin moments in the form of spinors allows us to treat several subtleties due to the geometrical nature of polarization. A clear interpretation of the polarized mixing distortions arises, as we show that a rotation of the spinor with frequency is naturally induced from the distribution of spectral parameters and polarization angles. We demonstrate that, no general pivot can be defined ensuring the vanishing of the first order in the non-perturbative regime. In the perturbative regime, however, such a pivot can be found and hides interesting physics. Correcting for this pivot naturally gives rise, in addition to a shift of the spectral parameters, to a frequency-dependent rotation of the polarization angle of predictable spectral dependence. It can also bring extra modulations to the polarized intensity, for example in the case of gray-bodies.

We explored scenarios of increasing complexity, considering several canonical SEDs examples of first importance for astrophysics. Even when dealing with highly complex signals along the line of sight, we showed that the use of spin moments allows us to model the distorted polarized intensity and the frequency dependent rotation of the polarization angle, including only a few terms in the expansion.

We also discussed the effect of more complex averaging processes such as spherical harmonics mixing and instrumental mixing as well as non trivial polarization specific situations like SED variations at the voxel-level and Faraday rotation effect. The spin-moment formalism still applies in all these scenarios. In these increasingly complex cases, however, the interpretation of the moment coefficients becomes blurry and the various mixing processes cannot be expected to be properly disentangled. Doing so, we also rederived formally the expressions used in previous works, treating $B$-mode signal as an intensity at the map \citep{Remazeilles_etal_2016} and power-spectra levels \citep{Azzoni,Vacher21}.

This work opens the door to several follow-up applications. On the theoretical side, several discussions remain open on the details of the generalization of the formalism at the power-spectrum level, especially for $E$- and $B$-mode applications. One could also think of links with cosmic birefringence, which recently received increased attention \citep[see e.g.][]{Diego2022BiRe}. More broadly, a lot of room is left for application of the spin moments as we defined them. One can think first to component separation where the spin moments could be competitive to model foreground distortions at the map level on large scales where a lot of averaging is done, calling for a comparative study with other pixel-based methods of component separation. Interesting questions related to Galactic physics have also to be addressed with the spin moments, such as the possibility to disentangle dust composition and Galactic magnetic field effects or tackle the Faraday rotation. One can also think of applications to SZ effect or spectral distortions, all topics that are left for future explorations.

\vspace{5mm}
\noindent
{\small
{\it Acknowledgments:}
%-------------------
LV would like to thanks François Boulanger for numerous and fruitful discussions.

JC was supported by the Royal Society as a Royal Society University Research Fellow at the University of Manchester, UK (No.~URF/R/191023).
This work was also supported by the ERC Consolidator Grant {\it CMBSPEC} (No.~725456) as part of the European Union's Horizon 2020 research and innovation program.
}

\bibliographystyle{aa}
\bibliography{bi}

\begin{thebibliography}{54}
\expandafter\ifx\csname natexlab\endcsname\relax\def\natexlab#1{#1}\fi

\bibitem[{{Aiola} {et~al.}(2020){Aiola}, {Calabrese}, {Maurin}, {Naess},
  {Schmitt}, {Abitbol}, {Addison}, {Ade}, {Alonso}, {Amiri}, {Amodeo},
  {Angile}, {Austermann}, {Baildon}, {Battaglia}, {Beall}, {Bean}, {Becker},
  {Bond}, {Bruno}, {Calafut}, {Campusano}, {Carrero}, {Chesmore}, {Cho},
  {Choi}, {Clark}, {Cothard}, {Crichton}, {Crowley}, {Darwish}, {Datta},
  {Denison}, {Devlin}, {Duell}, {Duff}, {Duivenvoorden}, {Dunkley},
  {D{\"u}nner}, {Essinger-Hileman}, {Fankhanel}, {Ferraro}, {Fox}, {Fuzia},
  {Gallardo}, {Gluscevic}, {Golec}, {Grace}, {Gralla}, {Guan}, {Hall},
  {Halpern}, {Han}, {Hargrave}, {Hasselfield}, {Helton}, {Henderson},
  {Hensley}, {Hill}, {Hilton}, {Hilton}, {Hincks}, {Hlo{\v{z}}ek}, {Ho},
  {Hubmayr}, {Huffenberger}, {Hughes}, {Infante}, {Irwin}, {Jackson}, {Klein},
  {Knowles}, {Koopman}, {Kosowsky}, {Lakey}, {Li}, {Li}, {Li}, {Lokken},
  {Louis}, {Lungu}, {MacInnis}, {Madhavacheril}, {Maldonado}, {Mallaby-Kay},
  {Marsden}, {McMahon}, {Menanteau}, {Moodley}, {Morton}, {Namikawa}, {Nati},
  {Newburgh}, {Nibarger}, {Nicola}, {Niemack}, {Nolta}, {Orlowski-Sherer},
  {Page}, {Pappas}, {Partridge}, {Phakathi}, {Pisano}, {Prince}, {Puddu}, {Qu},
  {Rivera}, {Robertson}, {Rojas}, {Salatino}, {Schaan}, {Schillaci}, {Sehgal},
  {Sherwin}, {Sierra}, {Sievers}, {Sifon}, {Sikhosana}, {Simon}, {Spergel},
  {Staggs}, {Stevens}, {Storer}, {Sunder}, {Switzer}, {Thorne}, {Thornton},
  {Trac}, {Treu}, {Tucker}, {Vale}, {Van Engelen}, {Van Lanen}, {Vavagiakis},
  {Wagoner}, {Wang}, {Ward}, {Wollack}, {Xu}, {Zago}, \& {Zhu}}]{ACT}
{Aiola}, S., {Calabrese}, E., {Maurin}, L., {et~al.} 2020, \jcap, 2020, 047

\bibitem[{Azzoni {et~al.}(2021)Azzoni, Abitbol, Alonso, Gough, Katayama, \&
  Matsumura}]{Azzoni}
Azzoni, S., Abitbol, M., Alonso, D., {et~al.} 2021, Journal of Cosmology and
  Astroparticle Physics, 2021, 047

\bibitem[{{Bennett et al}(2013)}]{WMAPfg}
{Bennett et al}, C.~L. 2013, \apjs, 208, 20

\bibitem[{{Brout} {et~al.}(1978){Brout}, {Englert}, \&
  {Gunzig}}]{inflationhist1}
{Brout}, R., {Englert}, F., \& {Gunzig}, E. 1978, Annals of Physics, 115, 78

\bibitem[{{Carlstrom} {et~al.}(2002){Carlstrom}, {Holder}, \&
  {Reese}}]{CosmoSZ}
{Carlstrom}, J.~E., {Holder}, G.~P., \& {Reese}, E.~D. 2002, \araa, 40, 643

\bibitem[{{Chluba} {et~al.}(2021){Chluba}, {Abitbol}, {Aghanim},
  {Ali-Ha{\"\i}moud}, {Alvarez}, {Basu}, {Bolliet}, {Burigana}, {de Bernardis},
  {Delabrouille}, {Dimastrogiovanni}, {Finelli}, {Fixsen}, {Hart},
  {Hern{\'a}ndez-Monteagudo}, {Hill}, {Kogut}, {Kohri}, {Lesgourgues},
  {Maffei}, {Mather}, {Mukherjee}, {Patil}, {Ravenni}, {Remazeilles}, {Rotti},
  {Rubi{\~n}o-Martin}, {Silk}, {Sunyaev}, \& {Switzer}}]{ChlubaVoy2050}
{Chluba}, J., {Abitbol}, M.~H., {Aghanim}, N., {et~al.} 2021, Experimental
  Astronomy, 51, 1515

\bibitem[{{Chluba} {et~al.}(2015){Chluba}, {Dai}, {Grin}, {Amin}, \&
  {Kamionkowski}}]{Chluba2015tensors}
{Chluba}, J., {Dai}, L., {Grin}, D., {Amin}, M.~A., \& {Kamionkowski}, M. 2015,
  \mnras, 446, 2871

\bibitem[{{Chluba} {et~al.}(2017){Chluba}, {Hill}, \& {Abitbol}}]{Chluba}
{Chluba}, J., {Hill}, J.~C., \& {Abitbol}, M.~H. 2017, \mnras, 472, 1195

\bibitem[{{Chluba} \& {Sunyaev}(2004)}]{Chluba2004SB}
{Chluba}, J. \& {Sunyaev}, R.~A. 2004, \aap, 424, 389

\bibitem[{{Chluba} {et~al.}(2013){Chluba}, {Switzer}, {Nelson}, \&
  {Nagai}}]{Chluba2012SZpack}
{Chluba}, J., {Switzer}, E., {Nelson}, K., \& {Nagai}, D. 2013, \mnras, 430,
  3054

\bibitem[{{CMB-S4 Collaboration}(2019)}]{CMBS4}
{CMB-S4 Collaboration}. 2019, arXiv e-prints, arXiv:1907.04473

\bibitem[{Delabrouille {et~al.}(2021)Delabrouille, Abitbol, Aghanim,
  Ali-Haïmoud, Alonso, Alvarez, Banday, Bartlett, Baselmans, Basu, Battaglia,
  Bermejo-Climent, Bernal, Bethermin, Bolliet, Bonato, Bouchet, Breysse,
  Burigana, \& Zubeldia}]{Voyage2050}
Delabrouille, J., Abitbol, M., Aghanim, N., {et~al.} 2021, Experimental
  Astronomy, 51

\bibitem[{{Diego-Palazuelos} {et~al.}(2022){Diego-Palazuelos}, {Eskilt},
  {Minami}, {Tristram}, {Sullivan}, {Banday}, {Barreiro}, {Eriksen},
  {G{\'o}rski}, {Keskitalo}, {Komatsu}, {Mart{\'\i}nez-Gonz{\'a}lez}, {Scott},
  {Vielva}, \& {Wehus}}]{Diego2022BiRe}
{Diego-Palazuelos}, P., {Eskilt}, J.~R., {Minami}, Y., {et~al.} 2022, \prl,
  128, 091302

\bibitem[{Draine \& Fraisse(2009)}]{Draine_2009}
Draine, B.~T. \& Fraisse, A.~A. 2009, The Astrophysical Journal, 696, 1

\bibitem[{{Ferri{\`e}re}(2001)}]{Ferriere}
{Ferri{\`e}re}, K.~M. 2001, Reviews of Modern Physics, 73, 1031

\bibitem[{{Goldberg} {et~al.}(1967){Goldberg}, {Macfarlane}, {Newman},
  {Rohrlich}, \& {Sudarshan}}]{Goldberg1967}
{Goldberg}, J.~N., {Macfarlane}, A.~J., {Newman}, E.~T., {Rohrlich}, F., \&
  {Sudarshan}, E.~C.~G. 1967, Journal of Mathematical Physics, 8, 2155

\bibitem[{{Guth}(1981)}]{inflationhist3}
{Guth}, A.~H. 1981, \prd, 23, 347

\bibitem[{{Heald}(2015)}]{Heald-FR}
{Heald}, G. 2015, in Astrophysics and Space Science Library, Vol. 407, Magnetic
  Fields in Diffuse Media, ed. A.~{Lazarian}, E.~M. {de Gouveia Dal Pino}, \&
  C.~{Melioli}, 41

\bibitem[{Hoseinpour {et~al.}(2020)Hoseinpour, Zarei, Orlando, Bartolo, \&
  Matarrese}]{CMBVmodes}
Hoseinpour, A., Zarei, M., Orlando, G., Bartolo, N., \& Matarrese, S. 2020,
  Phys. Rev. D, 102, 063501

\bibitem[{{Hutton} {et~al.}(2015){Hutton}, {Ferreras}, \&
  {Yershov}}]{vardustdisk}
{Hutton}, S., {Ferreras}, I., \& {Yershov}, V. 2015, \mnras, 452, 1412

\bibitem[{{Ichiki} {et~al.}(2019){Ichiki}, {Kanai}, {Katayama}, \&
  {Komatsu}}]{moment_polar}
{Ichiki}, K., {Kanai}, H., {Katayama}, N., \& {Komatsu}, E. 2019, Progress of
  Theoretical and Experimental Physics, 2019, 033E01

\bibitem[{{Inomata} \& {Kamionkowski}(2019)}]{Inomata2019Circ}
{Inomata}, K. \& {Kamionkowski}, M. 2019, \prd, 99, 043501

\bibitem[{{Jaffe} {et~al.}(2013){Jaffe}, {Ferri{\`e}re}, {Banday}, {Strong},
  {Orlando}, {Mac{\'\i}as-P{\'e}rez}, {Fauvet}, {Combet}, \&
  {Falgarone}}]{Jaffe2013}
{Jaffe}, T.~R., {Ferri{\`e}re}, K.~M., {Banday}, A.~J., {et~al.} 2013, \mnras,
  431, 683

\bibitem[{{Kamionkowski} {et~al.}(1997){Kamionkowski}, {Kosowsky}, \&
  {Stebbins}}]{Kamionkowski}
{Kamionkowski}, M., {Kosowsky}, A., \& {Stebbins}, A. 1997, \prd, 55, 7368

\bibitem[{{Kogut} {et~al.}(2011){Kogut}, {Fixsen}, {Chuss}, {Dotson}, {Dwek},
  {Halpern}, {Hinshaw}, {Meyer}, {Moseley}, {Seiffert}, {Spergel}, \&
  {Wollack}}]{PIXIE}
{Kogut}, A., {Fixsen}, D.~J., {Chuss}, D.~T., {et~al.} 2011, \jcap, 2011, 025

\bibitem[{{LiteBIRD Collaboration}(2022)}]{Ptep}
{LiteBIRD Collaboration}. 2022, in PTEP, Vol. 11443, PTEP, 114432F

\bibitem[{{Mangilli} {et~al.}(2021){Mangilli}, {Aumont}, {Rotti}, {Boulanger},
  {Chluba}, {Ghosh}, \& {Montier}}]{Mangilli}
{Mangilli}, A., {Aumont}, J., {Rotti}, A., {et~al.} 2021, \aap, 647, A52

\bibitem[{{Montero-Camacho} \& {Hirata}(2018)}]{Hirata2018Circ}
{Montero-Camacho}, P. \& {Hirata}, C.~M. 2018, \jcap, 2018, 040

\bibitem[{{Mroczkowski} {et~al.}(2019){Mroczkowski}, {Nagai}, {Basu}, {Chluba},
  {Sayers}, {Adam}, {Churazov}, {Crites}, {Di Mascolo}, {Eckert},
  {Macias-Perez}, {Mayet}, {Perotto}, {Pointecouteau}, {Romero}, {Ruppin},
  {Scannapieco}, \& {ZuHone}}]{Mroczkowski2019}
{Mroczkowski}, T., {Nagai}, D., {Basu}, K., {et~al.} 2019, \ssr, 215, 17

\bibitem[{{Newman} \& {Penrose}(1966)}]{NewmannPenrose1966}
{Newman}, E.~T. \& {Penrose}, R. 1966, Journal of Mathematical Physics, 7, 863

\bibitem[{{Newville} {et~al.}(2016){Newville}, {Stensitzki}, {Allen}, {Rawlik},
  {Ingargiola}, \& {Nelson}}]{Lmfit}
{Newville}, M., {Stensitzki}, T., {Allen}, D.~B., {et~al.} 2016, {Lmfit:
  Non-Linear Least-Square Minimization and Curve-Fitting for Python}

\bibitem[{{Pelgrims} {et~al.}(2021){Pelgrims}, {Clark}, {Hensley},
  {Panopoulou}, {Pavlidou}, {Tassis}, {Eriksen}, \& {Wehus}}]{pelgrims2021}
{Pelgrims}, V., {Clark}, S.~E., {Hensley}, B.~S., {et~al.} 2021, \aap, 647, A16

\bibitem[{{PICO Collaboration}(2019)}]{PICO}
{PICO Collaboration}. 2019, in \baas, Vol.~51, 194

\bibitem[{{Planck Collaboration}(2014{\natexlab{a}})}]{PlanckIX-2013}
{Planck Collaboration}. 2014{\natexlab{a}}, \aap, 571, A9

\bibitem[{{Planck Collaboration}(2014{\natexlab{b}})}]{Planck2014dust}
{Planck Collaboration}. 2014{\natexlab{b}}, \aap, 566, A55

\bibitem[{{Planck Collaboration}(2015)}]{Planck2015dust}
{Planck Collaboration}. 2015, \aap, 576, A107

\bibitem[{{Planck Collaboration}(2016)}]{planck_2015_overview}
{Planck Collaboration}. 2016, \aap, 594, A1

\bibitem[{{Planck Collaboration}(2017)}]{PlanckL}
{Planck Collaboration}. 2017, \aap, 599, A51

\bibitem[{{Planck Collaboration}(2020{\natexlab{a}})}]{PlanckCompoSep}
{Planck Collaboration}. 2020{\natexlab{a}}, \aap, 641, A4

\bibitem[{{Planck Collaboration}(2020{\natexlab{b}})}]{PlanckDust2}
{Planck Collaboration}. 2020{\natexlab{b}}, \aap, 641, A11

\bibitem[{{Remazeilles} {et~al.}(2016){Remazeilles}, {Dickinson}, {Eriksen}, \&
  {Wehus}}]{Remazeilles_etal_2016}
{Remazeilles}, M., {Dickinson}, C., {Eriksen}, H.~K.~K., \& {Wehus}, I.~K.
  2016, \mnras, 458, 2032

\bibitem[{{Remazeilles} {et~al.}(2021){Remazeilles}, {Rotti}, \&
  {Chluba}}]{RemazeillesmomentsILC}
{Remazeilles}, M., {Rotti}, A., \& {Chluba}, J. 2021, \mnras, 503, 2478

\bibitem[{{Rotti} \& {Chluba}(2021)}]{MILC}
{Rotti}, A. \& {Chluba}, J. 2021, \mnras, 500, 976

\bibitem[{{Sayre} {et~al.}(2020){Sayre}, {Reichardt}, {Henning}, {Ade},
  {Anderson}, {Austermann}, {Avva}, {Beall}, {Bender}, {Benson}, {Bianchini},
  {Bleem}, {Carlstrom}, {Chang}, {Chaubal}, {Chiang}, {Citron}, {Corbett
  Moran}, {Crawford}, {Crites}, {de Haan}, {Dobbs}, {Everett}, {Gallicchio},
  {George}, {Gilbert}, {Gupta}, {Halverson}, {Harrington}, {Hilton}, {Holder},
  {Holzapfel}, {Hrubes}, {Huang}, {Hubmayr}, {Irwin}, {Knox}, {Lee}, {Li},
  {Lowitz}, {McMahon}, {Meyer}, {Mocanu}, {Montgomery}, {Nadolski}, {Natoli},
  {Nibarger}, {Noble}, {Novosad}, {Padin}, {Patil}, {Pryke}, {Ruhl},
  {Saliwanchik}, {Schaffer}, {Sievers}, {Smecher}, {Stark}, {Tucker},
  {Vanderlinde}, {Veach}, {Vieira}, {Wang}, {Whitehorn}, {Wu}, {Yefremenko}, \&
  {SPTpol Collaboration}}]{SPT}
{Sayre}, J.~T., {Reichardt}, C.~L., {Henning}, J.~W., {et~al.} 2020, \prd, 101,
  122003

\bibitem[{{Schlafly} {et~al.}(2016){Schlafly}, {Meisner}, {Stutz},
  {Kainulainen}, {Peek}, {Tchernyshyov}, {Rix}, {Finkbeiner}, {Covey}, {Green},
  {Bell}, {Burgett}, {Chambers}, {Draper}, {Flewelling}, {Hodapp}, {Kaiser},
  {Magnier}, {Martin}, {Metcalfe}, {Wainscoat}, \& {Waters}}]{vardustdisk2}
{Schlafly}, E.~F., {Meisner}, A.~M., {Stutz}, A.~M., {et~al.} 2016, \apj, 821,
  78

\bibitem[{{Starobinsky}(1980)}]{inflationhist2}
{Starobinsky}, A.~A. 1980, Physics Letters B, 91, 99

\bibitem[{Stompor \& Efstathiou(1999)}]{Lensing}
Stompor, R. \& Efstathiou, G. 1999, Monthly Notices of the Royal Astronomical
  Society, 302, 735

\bibitem[{{Tassis} \& {Pavlidou}(2015)}]{tassis}
{Tassis}, K. \& {Pavlidou}, V. 2015, \mnras, 451, L90

\bibitem[{{The Simons Observatory collaboration}(2019)}]{SimonsObservatory}
{The Simons Observatory collaboration}. 2019, in \baas, Vol.~51, 147

\bibitem[{{Thorne} {et~al.}(2017){Thorne}, {Dunkley}, {Alonso}, \&
  {N{\ae}ss}}]{Pysm}
{Thorne}, B., {Dunkley}, J., {Alonso}, D., \& {N{\ae}ss}, S. 2017, \mnras, 469,
  2821

\bibitem[{{Vacher} {et~al.}(2022){Vacher}, {Aumont}, {Montier}, {Azzoni},
  {Boulanger}, \& {Remazeilles}}]{Vacher21}
{Vacher}, L., {Aumont}, J., {Montier}, L., {et~al.} 2022, \aap, 660, A111

\bibitem[{Wise(2019)}]{ReioWise}
Wise, J.~H. 2019, Contemporary Physics, 60, 145

\bibitem[{{Ysard} {et~al.}(2013){Ysard}, {Abergel}, {Ristorcelli}, {Juvela},
  {Pagani}, {K{\"o}nyves}, {Spencer}, {White}, \& {Zavagno}}]{Ysard2013}
{Ysard}, N., {Abergel}, A., {Ristorcelli}, I., {et~al.} 2013, \aap, 559, A133

\bibitem[{{Zaldarriaga} \& {Seljak}(1997)}]{Zaldarriaga1997}
{Zaldarriaga}, M. \& {Seljak}, U. 1997, \prd, 55, 1830

\end{thebibliography}

\begin{appendix}
\section{Alternative approaches}
\label{sec:normalization}
%------------------------------------
For astrophysical applications, it is common to normalize the modified blackbody SED in every pixel as in \cite{Pysm}
%------------------------------------
\begin{align}
    \hatpol^{\rm mBB}_\nu(\vecn)= \frac{\hatpol^{\rm mBB}_\nu(T(\vecn),\beta(\vecn))}{\hatpol^{\rm mBB}_{\nu_0}(T(\vecn),\beta(\vecn))}.
\end{align}
%------------------------------------
Since here both $\beta$ and $T$ are treated as spatially varying parameters, this is no longer a constant SED renormalization once line-of-sight effects are included. This choice therefore complicated the expansion, leading to rescaling of the moments \citep[e.g., see Eq.(4) of][]{Vacher21}, without any physical meaning or insight being added. 
At the pixel level, this complication can be avoided by setting 
%------------------------------------
\begin{align}
    \hatpol^{\rm mBB*}_\nu(\vecn)= \frac{\hatpol^{\rm mBB}_\nu(T(\vecn),\beta(\vecn))}{\hatpol^{\rm mBB}_{\nu_0}(\bar{T}(\vecn),\bar{\beta}(\vecn))},
\end{align}
%------------------------------------
which simply takes out the leading order term in the expansion in Eq.~\eqref{eq:mBB_momentExp}. However, when extending to applications on the full sky (or when averaging pixels), pixel to pixel variations of $\bar{\beta}$ and $\bar{T}$ will come into play. In this case, one should better normalize using one constant $\bar{\beta}$ and $\bar{T}$ across the sky to avoid additional complications from the variation of the normalization. 

%Doing so, one should be careful. While the power-law factor simply leads to constants rescalings of the moments, the spectral derivatives with respect to $\beta_{\rm GB}$ would add extra un-physical terms in the spectral shapes of the expansion due to the choice of normalization. In principle they can easily be treated (see e.g. Eq.(4) of \cite{Vacher21}), but one would rather want to avoid this by normalizing the whole map by a single modified blackbody of constant spectral parameters $\beta_0$ and $T_0$.

\section{Choice of weights}
\label{sec:weights}
%------------------------------------
As discussed already in \citet{Chluba}, when deriving the moment expansion, one can always reexpress the SEDs with another choice of units\footnote{Some changes of units as e.g. Jy.sr$^{-1} \to \mu{\rm K}_{\rm CMB}$ are slightly subtler and introduce new frequency-dependent terms.}  or with respect to an alternative choice of spectral variables. Such a choice should be motivated by the physics and the numerical behavior of the problem considered. Ultimately, doing so will simply be equivalent to a re-scaling of the weight coefficients appearing in the moment expansion. Such a change can have a significant impact on the convergence rate of the expansion and the interpretations of the moments coefficients but is mathematically equivalent.

For example, in the case of the gray-body, one could equivalently define the new weights $A'(T)= A T^3 $ and expand around $T$ instead of $\beta_{\rm GB}$. The fundamental SED then reads
%--------------------------
\begin{equation}
\hatpol_\nu^{\rm GB} = A' \left(\frac{\bar{T}}{T}\right)^3
\frac{x^3}{
{\rm e}^{\frac{x\bar{T}}{T}}-1}.
\end{equation}
%--------------------------
The moment expansion will change accordingly as displayed in Eq.~(43) of \citet{Chluba}. This allows us to interpret the spin moments and pivot directly in term of temperatures: %--------------------------------------
\begin{subequations}
\begin{align}
    \bar{A'} &= \langle A' \rangle = \langle A T^3 \rangle\\
    \bar{T}
    &= \frac{\langle A'T\rangle}{\bar{A'}} = \frac{\langle AT^4\rangle}{\langle A T^3 \rangle} \\
    \mathcal{W}_\alpha^{T} 
    &=\frac{\langle A'\,(T-\bar{T})^\alpha \expf{2\i\gamma}\rangle}{\bar{A'}} = \frac{\langle AT^3\,(T-\bar{T})^\alpha \expf{2\i\gamma}\rangle}{\langle A T^3 \rangle}.
\end{align}
\end{subequations}
The exact same reasoning applies to a change of the power law's reference frequency $\nu_0 \to \nu'_0$ with the simple reweighting $A'=A\left(\nu'_0/\nu_0\right)^\beta$. However, in this case, no new SED derivatives are created, so the two expansions are identical.
\end{appendix}
\end{document}